\documentclass{ptephy_v1}

\usepackage{graphicx}
\usepackage{url}
\usepackage{bm}
\usepackage{amssymb}	
\usepackage{amsthm}	
\usepackage[sc]{mathpazo}
\usepackage{mathrsfs}
\usepackage{booktabs}
\usepackage{rotating}
\usepackage[flushleft]{threeparttable}
\usepackage{amsmath}
\makeatletter
\renewcommand*\env@matrix[1][*\c@MaxMatrixCols c]{%
  \hskip -\arraycolsep
  \let\@ifnextchar\new@ifnextchar
  \array{#1}}
\makeatother

\newcommand{\beq}{\begin{equation}}
\newcommand{\eeq}{\end{equation}}

\newcommand{\ud}{\mathrm{d}}
\newcommand{\ui}{\mathrm{i}}
\newcommand{\uk}{\mathrm{k}}
\newcommand{\um}{\mathrm{m}}
\newcommand{\un}{\mathrm{n}}
\newcommand{\uJ}{\mathrm{J}}
\newcommand{\Tr}{\mathrm{Tr}}

\newcommand{\Bc}{\mathcal{B}}

\newcommand{\Ic}{\mathcal{I}}

\begin{document}

\title{Modification of  relativistic beam fields under the influence of external conducting
 and ferromagnetic flat boundaries}

\author{B.~B.~Levchenko}%
\affil{D.V. Skobeltsyn Institute of Nuclear Physics, M.V. Lomonosov Moscow State University, 
119991 Moscow, Russian Federation \email{levtchen@mail.desy.de}}

\begin{abstract}%
We derive  analytical expressions for external fields of a  relativistic 
bunch of charged particles with a circular and an elliptical cross section
 under different boundary conditions and interaction of the fields with an accelerator structural elements. 
The particle density in the bunch is assumed to be  uniform as well as non-uniform.
At distances far apart from the bunch, in free space the field reduces to the relativistic modified Coulomb 
form for a pointlike charge and at small distances the expressions reproduce the external 
fields of a continuous beam. 
In an ultra-relativistic limit the longitudinal components of the internal and external 
electric fields of the  bunch are strongly suppressed by the Lorentz factor.
If the bunch is  surrounded by conducting surfaces, the bunch self-fields are modified. 
Image fields generated by a bunch between two parallel conducting plates are studied in detail. 
Exact summation of the electric, $E_y$,  and magnetic, $B_x$, image field components 
allows  the infinite series to be represented in terms of elementary trigonometric functions.
 The new expressions for modified fields are applied  to study 
image forces acting on the bunch constituents and the bunch as a whole.
The coherent and incoherent tune shifts for an arbitrary bunch displacement 
from the midplane are calculated in the framework of an improved linear theory,
 for both infinite and finite parallel flat surfaces. 
Moreover, the developed method allows us to  generalize  the Laslett image 
coefficients $\epsilon_1$, $\epsilon_2$, $\xi_1$, $\xi_2$ to the case of 
an arbitrary bunch offset and reveal relationships between these coefficients. 

Appendix \ref{appC} provides a brief historical background of the development of the 
method of electrical images.
\end{abstract}

\subjectindex{electrodynamics, exact solution, image field, Kelvin, Laslett coefficients, beam dynamics, tunes}

\maketitle

\section{\label{intro}Introduction}

In an accelerator, the charged beam is influenced by the environment matter (a beam
pipe, accelerator gaps, magnets, collimators, etc.), and a high-intensity
bunch of particles induces surface charges or currents into this environment. This
modifies the electric and magnetic fields around the bunch. There is a
relatively simple method to account for the effect of the environment
by introducing image charges and currents.
The mathematical technique of electrical images was developed  by W. Thomson 
(Lord Kelvin) \cite{Thomson:1845, Thomson_Rep_BA_1847, Thomson:1884}.  
The method of images has found application in various branches of physics, 
in particular, in hydrodynamics \cite{Stokes_Rep_BA_1847, Lamb:1932, Newman:1999}.
 Appendix \ref{appC} provides a brief historical background on 
the subject\footnote{ 20th century textbooks on classical 
electrodynamics do not specify  the author of the method of electrical images.}.

	Over fifty years ago Laslett \cite{Laslett:1963jn, Laslett:1987jn}   
analyzed the influence of the transverse 
space-charge phenomena,  due to image forces, on the instability of the coherent transverse 
motion of an intense beam.  
Methods of image field summation are described in his paper \cite{Laslett:1963jn}, 
which presented  some field  coefficients calculated for infinite parallel plate vacuum chambers, magnet poles, and vacuum chambers with elliptical cross sections and variable aspect ratios. 
The resulting  image fields were calculated only in the linear approximation and depend linearly on the deviations $\bar{y}$ and $y$  of the bunch center and the position of a test particle, respectively, from the axis (see Fig. 4). 
They act therefore like a quadrupole causing a coherent tune shift.
The approximation  used  is incorrect if the field observation point  $y$ is located far from the bunch or if  the bunch center $\bar{y}$ is close to a conducting wall.
 
In the present paper we consider this classical  problem  summation  of fields of images once again for a very simple geometry, namely,  an ultra-relativistic bunch moving between infinitely wide  parallel perfectly conducting plates. 
The problem is far from being  purely academic.  
In applications, in particular, in the study of dynamics of photoelectrons in the beam transport system \cite{Cimino:2004cv} and the electron cloud effect 
intensified by electron field emission in the flat collimator \cite{Levchenko:2006tm}, 
it is important to know the distribution of electromagnetic fields not only in the vicinity of  
the bunch, but in  the whole collimator gap.
 We have not found publications with attempts to sum up the series (\ref{s4-1}) (see Sect. \ref{imch}) 
in an approximation beyond the linear one. 
In Sects. \ref{imch} and \ref{imcu} we present  exact 1D solutions of the problem for electric 
and magnetic image fields. 
The preliminary results were presented in Ref. \cite{Levchenko:2010pri}.
1D solutions are canonically used in the calculation of tune 
shifts\footnote{Exact 2D solutions by the method of images and other applications will be presented elsewhere.}.

Before we solve the problem formulated above, in Sect. 2 we first derive  expressions  for the external
 electric and magnetic fields generated by a cylindrical and an elliptical 
 bunch of charged particles. 
The task is  specified  as follows.

The external radial electric field $\vec{E}_r$ and azimuthal magnetic induction $\vec{B}_{\phi}$
for a round unbunched relativistic beam of  radius $a$ and a uniform  charge density are described by 
\cite{Sands:1970ye,  Wiedemann:2007yf, Chao:1993zn}
\begin{eqnarray} 
 E_{r}&=&\kappa\frac{2q\lambda}{r},  \label{I-1}\\
 B_{\phi}&=&\frac{\mu_0}{4\pi}\frac{2q\lambda}{r}c\beta,\label{I-2}
\end{eqnarray}
where $\kappa= 1/4\pi\epsilon_0$,  $\lambda$ is the linear beam density,  $q$ is the 
charge,   $\beta = v/c$ is a normalized velocity of the beam constituents and $c$ the velocity of light \footnote{In Ref. \cite{Wiedemann:2007yf}, 
the azimuthal component $B_{\phi}$, Eqs. (18.51)-(18.52), 
includes the minus sign, in contradiction to Eq. (18.28) and
the expression of cylindrical coordinates of a vector field by Cartesian coordinates (Ref. \cite{BronSemen1973}, p. 630): $B_{\phi}=-B_{x}\sin\phi + B_{y}\cos\phi$.}. 
In many applications,  (\ref{I-1}) and (\ref{I-2}) are used to describe   fields of an individual bunch too. 
However, in the form (\ref{I-1}), (\ref{I-2}) the bunch fields  do not depend on the bunch energy and at large distances do not follow the Coulomb asymptotic.
This contrasts sharply with the fields produced (at  $t=0$) by a rapidly moving  single charge $q$
\beq\label{I-3}
  \vec{E}\,=\, \kappa\frac{q\,\gamma}{ r^2} \Big [ 
 \frac{1-\beta ^2}{1-\beta ^2\, sin^2\,\theta}\Big ]^{3/2}\,
\frac{\bf \vec{r}}{r},
\qquad
c\vec{B}\,\sim \vec \beta \times \vec E,
\eeq
where $\theta$ is the angle that the vector $ \vec{\bf r}$ makes with the $z$-axis. 
Along the direction of motion the electric field becomes weaker  
in $\gamma ^2$ times, while in the transverse direction the electric field is 
enhanced by the factor $\gamma$:
\beq
  E_r\,=\,  \kappa \frac{q\,\gamma }{r^2} .
\label{I-4}
\eeq
Here, $\gamma$ denotes the particle Lorentz factor.

This paper is organized as follows.
In the next section, we demonstrate how, when summing up the elementary electromagnetic fields generated by charged relativistic particles, effective external beam fields are formed.
 We derive  expressions for the transverse and longitudinal components of the bunch electric field, where the defects indicated above are rectified, and find the conditions at which the bunch fields are represented by Eqs. (\ref{I-1}) and (\ref{I-2}). 
Here we consider bunches shaped as a cylinder with a circular and an elliptical cross section. 
In Sect. 3 we discuss fields generated by a bunch with an arbitrary linear particle density and make a statement that in the ultra-relativistic limit $\gamma \rightarrow \infty $ the electric field takes  a universal  form. 

Sections 4 and 5 are devoted to the problem of finding  exact analytic expressions for the electric and magnet fields generated by a bunch moving between infinitely wide parallel conducting plates and magnet poles. 

In Sect. 6 we discuss image forces acting on the bunch constituents and the bunch as a whole and calculate in the framework of an improved linear approximation the coherent and incoherent tune shifts for an arbitrary bunch displacement from the midplane. 
In Sect 6.3, a practical example is considered as gradients of image fields in a finite-size collimator affecting the betatron frequency of the beam.
Conclusions are drawn in Sect. 7. 
Here we also compare results obtained by different authors with the use of  various techniques.
Detailed derivations of the obtained results are placed in  Appendixes \ref{appA} and  \ref{appB}.

\section{Self-fields of a charged  cylinder 
 with an elliptical cross section}

Let us consider a bunch of charged particles uniformly distributed with a  density $\rho$ within a cylinder of  length $L$ and  an elliptical  cross section.  
The ellipsoid semi-axes in the $x-y$ plane are $a$ and $b$ and the coordinate  $z$-axis is along the bunch axis.
Suppose that the bunch is moving along the z-axis with a relativistic velocity 
$\vec{v}=c\vec{\beta}$.

To compute the radial electric  field of such a rapidly moving  bunch, we have to sum up fields of the type  (\ref{I-3}), generated by the bunch  constituents. 
In this way we get \cite{Ferrario:2003}
\beq
  E_{\perp}(r,\xi,z)\,=\,\kappa\rho \gamma\big [zI_1\,+\,(L-z)I_2\big ]
\label{a2-1}
\eeq
with
\beq
I_1\,=\,\int \!\int\frac{[r-\sigma \cos(\xi-\phi)]\,\sigma\ud\sigma\ud\phi}
{[ r^2+\sigma^2-2r\sigma \cos(\xi-\phi)]
[\gamma^2 z^2+r^2+\sigma^2-2r\sigma \cos(\xi-\phi)]^{1/2}} 
\label{a2-i1}
\eeq
\beq
I_2\,=\,\int \!\int\frac{[r-\sigma \cos(\xi-\phi)]\,\sigma\ud\sigma\ud\phi }
{[ r^2+\sigma^2-2r\sigma \cos(\xi-\phi)]
[\gamma^2 (L-z)^2+r^2+\sigma^2-2r\sigma \cos(\xi-\phi)]^{1/2}}
\label{a2-2}
\eeq
where $\sigma$ is the distance in the $x-y$ plane from the $z$-axis  to the elementary charged volume and 
\beq
 0<\sigma \le \Sigma (\phi) =\frac{b}{\sqrt{1 - e^2\cos^2 \phi}},\ \ \ \ 0<\phi <2\pi,
\eeq
where $e=\sqrt{1-b^2/a^2}$ is the eccentricity of an ellipse and $a > b$.
Equation (\ref{a2-1}) represents the radial electric field at instant $t=0$ as observed at a distance $r$ from the bunch axis, at an angle $\xi$ relative to the $x$-axis and at a distance $z$ from the bunch tail.

In Ref.\cite{Ferrario:2003} integrals $I_1$ and $I_2$ were estimated only numerically, because the integrands were taken as they are.  
However,  the integrands  are easy to simplify if the bunch is ultra-relativistic, $\gamma \gg 1$, and   we would now like to calculate the field in vicinity   of the bunch but at distances  much larger than the bunch radius, $r \gg a$.

  To simplify this,  we make use of the notations
\begin{eqnarray}
 A &=& \sigma /r,\ \ B=A \cos(\xi-\phi),\ \ Y=A^2-2B,\nonumber \\
 C_1 &=& \Big[1+\gamma^2 z^2 /r^2\Big]^{-1},\ \ X=C_1\cdot Y \nonumber
\end{eqnarray}
and the integrand of $I_1$ can be written as
\beq
 (r^2+\gamma^2z^2)^{-1/2}A(1-B)(1+Y)^{-1}(1+X)^{-1/2}\,.
\label{a2-3}
\eeq
Now we expand the above expression in a power series by using $A$ as a small parameter and keeping only terms  up to the power $A^4$ at each step.

\subsection{Finite circular cylinder with a uniform particle density}
For a bunch shaped as a circular cylinder, $a=b$,  we may set $\xi=0$. 
Due to the fact that
\beq
 \int_0^{2\pi} \cos^{\rm 2k+1}\phi\,\ud\phi\,=\,0,
\eeq
all odd powers of $B$ vanish after integration over $\phi$. This greatly simplifies the series 
generated from Eq. (\ref{a2-3}). After lengthy algebraic manipulations  with Eq. (\ref{a2-3}), we get
\beq 
(r^2+\gamma^2z^2)^{-1/2}A\Big [1-(1+\frac{1}{2}C_1)A^2+(2+C_1+\frac{3}{2}C_1^2)B^2\Big ].
\label{a2-4}
\eeq
Substituting this expression in Eq. (\ref{a2-i1}), one gets
\beq
 I_1\,=\,\frac{\pi a^2}{r\sqrt{r^2+\gamma^2z^2}}\Big(1+\frac{3}{8}C_1^2\frac{a^2}{r^2}\Big ).
\label{s1-5}
\eeq
\begin{figure*}
\centering
\begin{minipage}[h]{.46\textwidth}
\includegraphics[width=\textwidth]{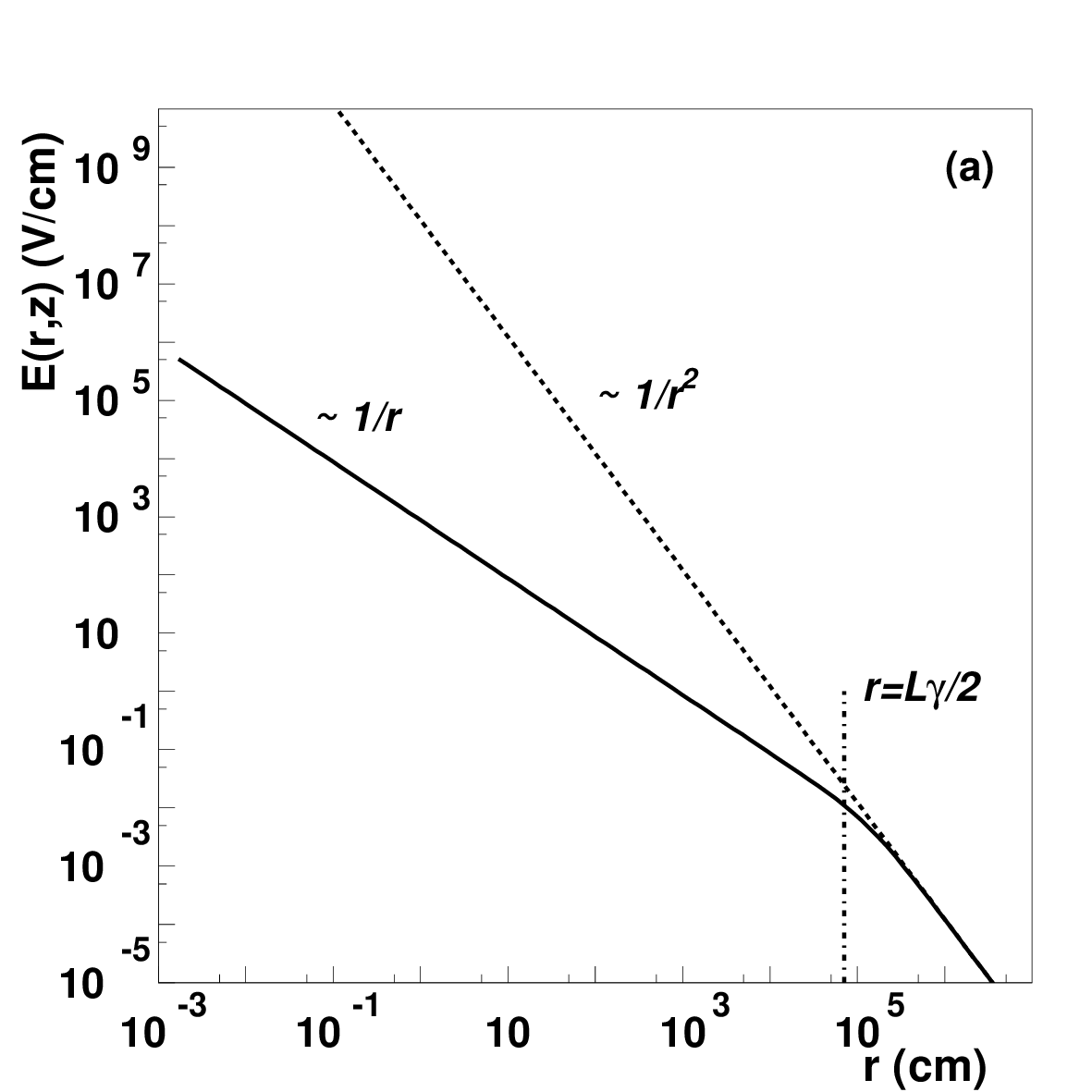}
\end{minipage}	\hspace*{+5mm}
\begin{minipage}[h]{.46\textwidth}
\includegraphics[width=\textwidth]{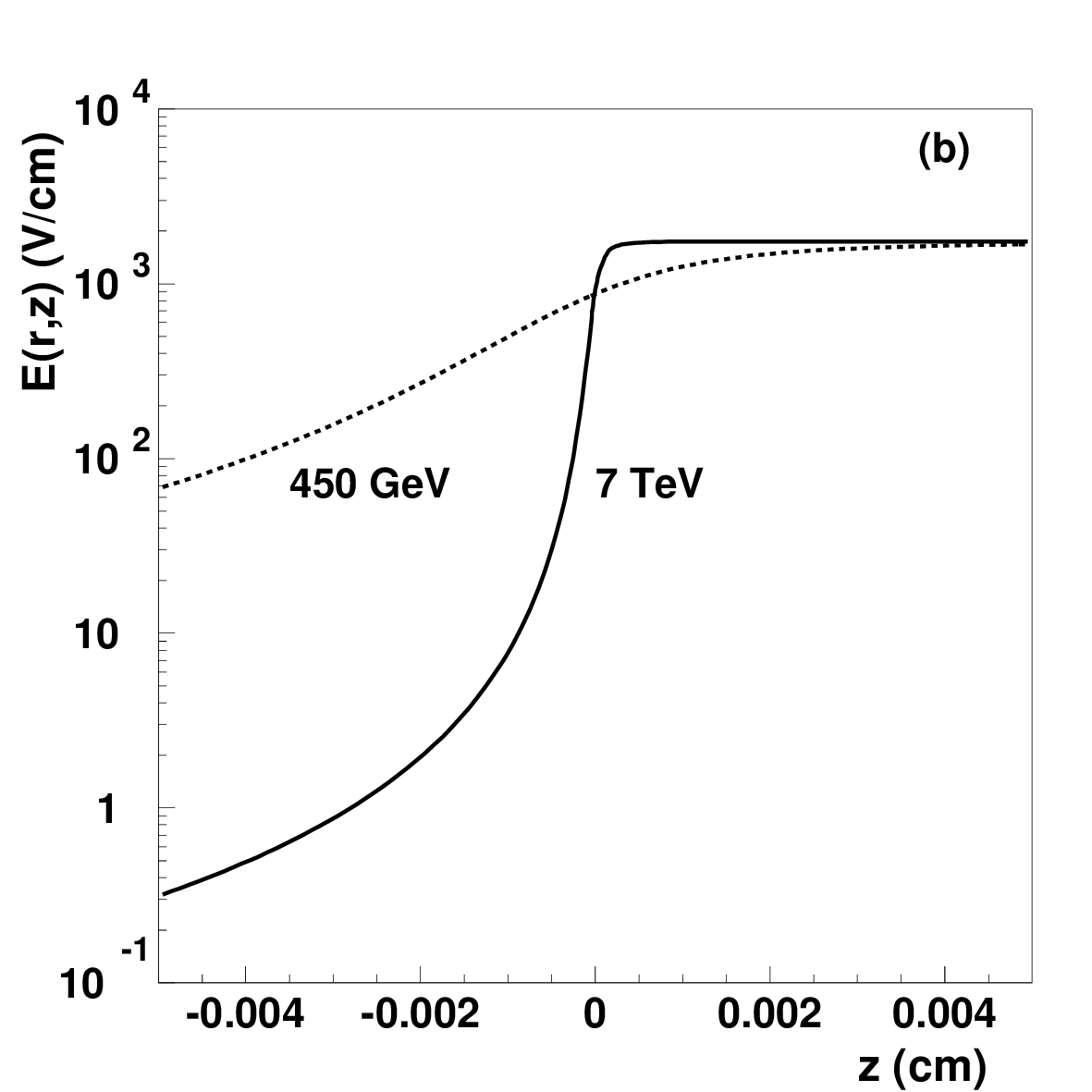}
\end{minipage}
\noindent
\caption{(a) The transverse profile of the electric field generated by a relativistic bunch of a circular cross section. 
The vertical dash-dotted line indicates the crossover point of the curve (\ref{I-4}) and (\ref{I-1}). 
Here $L=\sqrt{2\pi}\sigma_z$ and $\sigma_z$ denoting the r.m.s. bunch length.
(b) The radial field variation with $z$ near the bunch tail at fixed $r=1.0$ cm. 
For  comparison, the dashed line shows the field from a proton bunch at the energy 450 GeV. 
Calculations are performed with parameters corresponding to the LHC proton beam (see Table 1).}
\label{fig1} 
\end{figure*}
By changing $z^2$ to $(L-z)^2$ in (\ref{s1-5}), we obtain  for $I_2$ the following result
\beq
 I_2\,=\,\frac{\pi a^2}{r\sqrt{r^2+\gamma^2(L-z)^2}}\Big(1+\frac{3}{8}C_2^2\frac{a^2}{r^2}\Big),
\label{s1-6}
\eeq
where $C_2=\Big[1+\gamma^2(L-z)^2/r^2\Big]^{-1}$. 
Notice that for particles uniformly distributed   in the bunch volume, $\rho=qN_b/\pi a^2 n_b L=q\lambda/\pi a^2$, where $N_b$ is the total number of particles in the beam, $n_b$ the  number of bunches, 
and $\lambda$ the linear particle density.
Substituting  Eqs. (\ref{s1-5})$-$(\ref{s1-6}) in (\ref{a2-1}), finally we arrive at
\beq\label{s1-14}
E_{\perp}(r,z) =\kappa\frac{q\lambda\gamma}{r}\Big [\frac{z}
{\sqrt{r^2+\gamma^2z^2}}\Big(1 + \frac{3}{8}\frac{a^2}{r^2}C^2_1\Big) + 
\frac{L-z}{\sqrt{r^2+\gamma^2(L-z)^2}}\Big(1+ \frac{3}{8}\frac{a^2}{r^2}C^2_2\Big) \Big ].
\eeq
This equation describes the transverse component of the electric field produced by a rapidly moving circular bunch.
For $\gamma \gg	1$ the correction factors $C_1$ and $C_2$ can be neglected. 
In that case at the bunch surface, $r = a$, Eq. (\ref{s1-14}) exactly matchs the equation for the internal field \cite{Ferrario:2003}. 
Therefore, the condition $r\gg a$  used to derive Eq. (14) can be weakened and Eq. (\ref{s1-14}) is valid in the region $r \ge a$.

The field of a  relativistic bunch described by Eq. (\ref{s1-14})  has different behavior at distances far away from the bunch and for $r < L\gamma/2$. Figure~1(a) shows the radial field profile  as follows from (\ref{s1-14}). 
The parameters of the bunch  correspond to the nominal scenario of the LHC proton beam \cite{PDBook}.
At very large distances, $r \gg L\gamma/2$,   (\ref{s1-14}) reduces to the Coulomb form (\ref{I-4}) with $q$ replaced by $qN_b/n_b$. 
Calculations show that for a proton bunch at 7 TeV the Coulomb law is restored only at a distance of several kilometers from the bunch.
On the other hand, for $r \ll L\gamma/2$, Eq. (\ref{s1-14}) simplifies to a form  independent of the $z$-coordinate, which coincides with the external field Eq. (\ref{I-1}) of a continuous  beam with $\lambda=N_b/n_b L$.

The magnitude of the electric field varies drastically in the head and tail parts of the bunch.
For instance, in a very narrow transition region beyond the bunch tail, $z <0$, $|z|\ll r/\gamma$, the field strength decreases with $z$ as follows:
\beq
 E_{\perp}(r,z)\,\approx\,\kappa\frac{q\lambda}{r}\Big [ 1-\frac{r^2}{2\gamma^2(L-z)^2}\Big ].
\label{s1-15}
\eeq 
However, at larger $|z|$, the suppression of the radial field by the Lorentz factor becomes  dominant,
\beq
 E_{\perp}(r,z)\,\approx\,\kappa\frac{q\lambda r}{2\gamma^2}\, \frac{L(L-2z)}{z^2(L-z)^2}.
\label{s1-16}
\eeq 
This is shown in Fig.~1(b). The field strength decreases to more than three orders of magnitude at the distance of 4 $\mu$m beyond the bunch. One finds from Eq. (\ref{s1-14}) that at $z \ge L$ the field magnitude tends to zero in a similar way.

The longitudinal part of the electric field reaches the maximum value on the bunch axis, so that  
on axis the longitudinal electric field is given by \cite{Ferrario:2003}
\beq
E_z(0,z)\,=\,\kappa\frac{2\pi\rho}{\gamma}\{\sqrt{a^2+\gamma^2(L-z)^2}
-\sqrt{a^2+\gamma^2 z^2}+\gamma |z| -\gamma |L-z|\}.
\label{s1-17}
\eeq 
At $\gamma \gg 1$, the field magnitude outside the bunch is given by
\beq
E_z(0,z)\,=\,\kappa\frac{q\lambda}{\gamma^2}\Big [ \frac{1}{|L-z|} - \frac{1}{|z|}\Big ].
\label{s1-18}
\eeq 
Equation (\ref{s1-18}) shows that the longitudinal field is independent of the bunch radius and strongly  suppressed along the line of motion of the bunch.

The main result of this consideration is that due to features (\ref{s1-16}) and (\ref{s1-18}), 
the space-time distribution of the electric field around an ultra-relativistic circular 
bunch with a uniform particle density is well approximated by a step-like form:
\beq
 E(r,z,t)\,=\,\kappa\frac{2q\lambda}{r}\Big [\theta (z-\beta ct) -\theta (z-\beta ct -L) \Big ].
\label{s1-19}
\eeq 

Similarly, we can show that the azimuthal magnetic induction of the bunch is 
\beq
 B_{\phi}(r,z,t)\,=\, \frac{\mu_0}{4\pi}\frac{\beta c}{\kappa}E_{\perp}(r,z,t).
\label{s1-20}
\eeq 
%


\subsection{Finite elliptical cylinder with a uniform particle density}

Let us now consider a bunch shaped as an elliptical cylinder. 
We make use of the same notations as in the previous subsection. 
In the ultra-relativistic scenario, the correction factors $C_1$ and $C_2$ should be neglected from the beginning. By expanding the integrand of $I_1$ in a power series as above, we get
\begin{eqnarray}
&&(r^2+\gamma^2 z^2)^{-1/2} 
\Big \{
\sum_{\uk=0}^{\uk_{max}} A^{2\uk+1}\cos [2\uk(\xi-\phi)] \nonumber\\
&&+A^2 \cos(\xi-\phi)\big [1 -A^2(1-\cos [2(\xi-\phi)])     
+ A^4(1-2\cos [2(\xi-\phi)]+2\cos [4(\xi-\phi)])\nonumber\\
&&- A^6(1-2\cos [2(\xi-\phi)]+2\cos [4(\xi-\phi)] 
- 2\cos [6(\xi-\phi)] ) \big ] +\cdots \Big \}.
\label{s1-21}
\end{eqnarray}
%
 It can be proven that for an even function $f_{\rm n}(\phi)$
\beq 
\int_0^{2\pi}\ud\phi\cos^{2\uk+1}(\xi-\phi)
\int_0^{\Sigma (\phi)}\ud\sigma A^\un 
= \frac{b^{\un+1}}{(\un+1)r^\un}\int_0^{2\pi}f_\un(\phi)\cos^{2\uk+1}(\xi-\phi)\ud\phi = 0,
\label{s1-22}
\eeq 
where $\uk = 0,1, 2,\dots $. In our consideration
\beq
 f_\un(\phi)=(1-e^2\cos^2\phi )^{-(\un+1)/2}.
\label{s1-23}
\eeq 

The integral (\ref{a2-i1}) now can be solved with respect to $\sigma$ and $\phi$ by direct 
substitution of Eq. (\ref{s1-21}) and the use of Eq. (\ref{s1-22}): 
\beq
 I_1=\frac{\pi ab}{r\sqrt{r^2 + \gamma^2 z^2}}\frac{\sqrt{1-e^2}}{\pi}
\sum_{\uk=0}^{\uk_{max}}
\frac{b^{2\uk}}{(2k+2)r^{2\uk}}\cdot D_\uk\cdot \cos (2\uk\xi) .
\label{s1-24}
\eeq
The condition (\ref{s1-22}) causes all even terms in $A$ to vanish.
Here, 
\beq
 D_\uk=\int_0^{2\pi} \frac{\cos (2\uk\phi)\,\ud\phi}{(1-e^2\cos^2\phi )^{\uk+1}}\,=\,
d_\uk\cdot\frac{\pi e^{\rm 2k}}{(1-e^2)^{\uk+1/2}},
\eeq
with numerical coefficients $d_\uk=(2,1,3/4,5/8,35/64,$ $63/128,231/524,429/1024,...)$.
By changing $z^2$ to $(L- z)^2$ in Eq. (\ref{s1-24}), one finds $I_2$ too.

For particles uniformly distributed in the elliptical bunch volume, $\rho = q\lambda/\pi ab$ with
$\lambda = N_b/n_b L$ the linear particle density. Substituting equations for $I_1$ and $I_2$ in 
Eq. (\ref{a2-1}), finally we arrive at
\begin{eqnarray} 
E_{\perp}(r,z)&=&\kappa\frac{q\lambda\gamma}{r}
\Big (\frac{z}{\sqrt{r^2+\gamma^2z^2}}
+ \frac{L-z}{\sqrt{r^2+\gamma^2(L-z)^2}} \Big )   
\Big [1 + 
\frac{1}{4}\Big(\frac{ae}{r}\Big)^2\cos (2\xi) \nonumber \\ 
&+&\frac{1}{8}\Big(\frac{ae}{r}\Big)^4\cos (4\xi) 
+\!\!\frac{5}{64}\Big(\frac{ae}{r}\Big)^6\cos (6\xi) 
+\frac{7}{128}\Big(\frac{ae}{r}\Big)^8\cos (8\xi)
+\cdots   \Big ].
\label{s1-26}
\end{eqnarray}
This equation describes the transverse component of the electric field produced by a rapidly moving elliptical bunch of length $L$. 
The transverse part is modulated by an angular factor
that takes into account an  ellipticity of the bunch. The dimensions of the ellipse  enters only via  the ratio $ae/r$.

\begin{figure*}
\centering
\begin{minipage}[h]{.41\textwidth}
\vspace{+6mm}
\includegraphics[width=\textwidth]{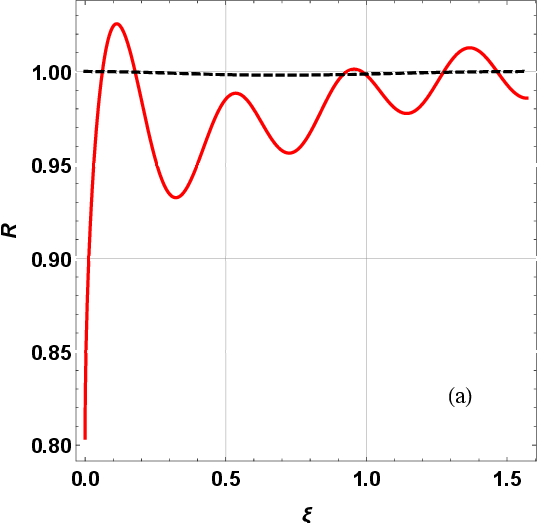}
\end{minipage}	\hspace*{+4mm}
\begin{minipage}[h]{.46\textwidth}
\includegraphics[width=\textwidth]{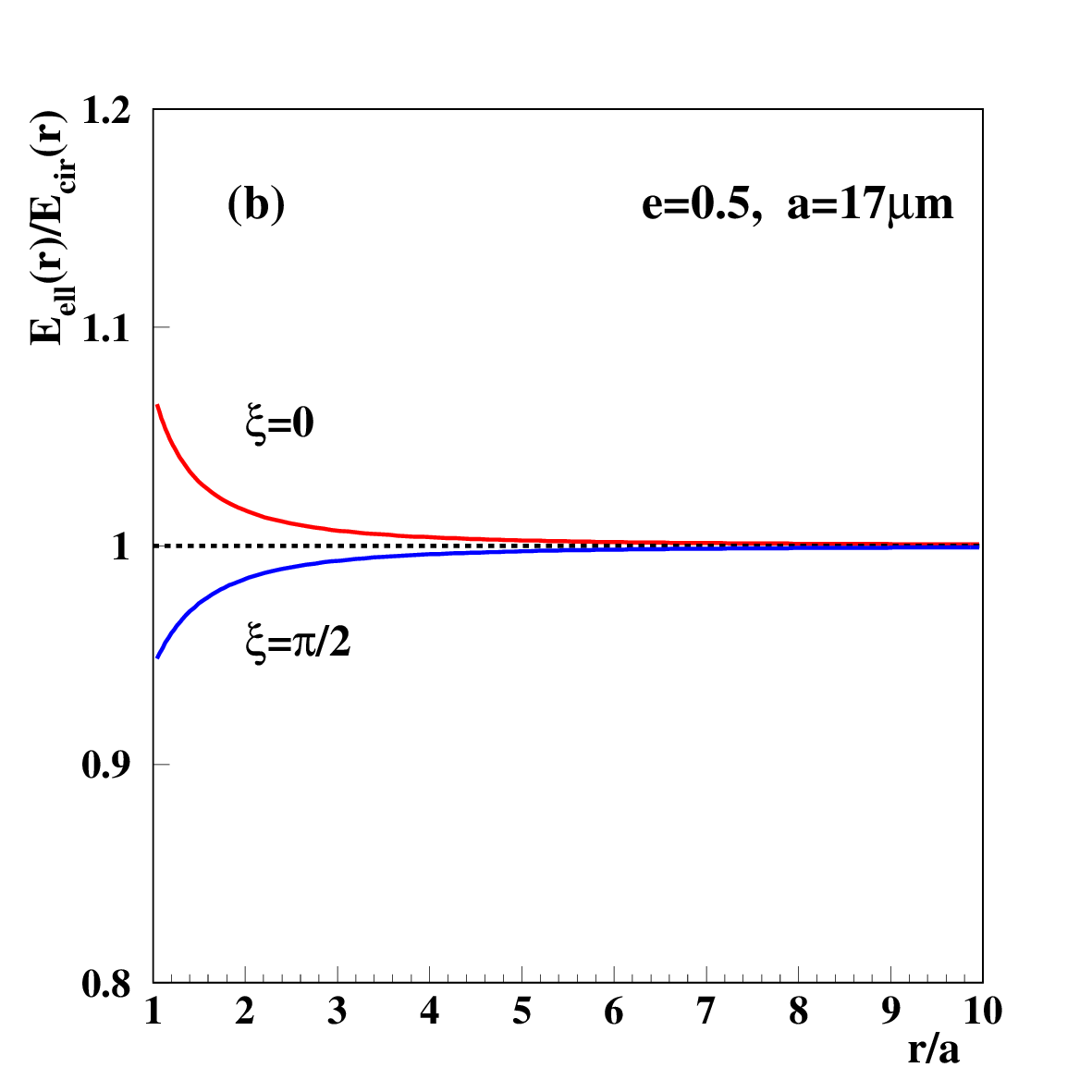}
\end{minipage}
\caption{(a) The ratio of electrical field (\ref{s1-27}) to  electrical field  (\ref{s1-30}) as a function of $\xi$ in the first quadrant. The calculation is done at $r/ae =1$ (full curve) and  $r/ae=2$ (dashed  curve) with $\uk_{\rm max}=7$ (see Eq. (\ref{s1-24})).
(b) The ratio of  the electrical field (\ref{s1-27}) created by an elliptical bunch ($e=0.5$ and $a=17 \mu$m)  to the electrical field (\ref{s1-19})  created by a  circular bunch 
($a=17 \mu$m) as a function of the radial distance 
at the azimuthal angle $\xi=0$ and $\xi=\pi/2$. }
\label{fig2} 
\end{figure*}

The arguments used in  deriving   Eq. (\ref{s1-19}) are also applicable here. 
The electric field of an ultra-relativistic elliptical bunch  is therefore well approximated by 
a step-like form:
\begin{eqnarray} 
E(r,z,\xi,t)&=&\kappa\frac{2q\lambda}{r}
\Big [\theta (z-\beta ct) -\theta (z-\beta ct -L) \Big ] 
\Big [1+
\frac{1}{4}\Big(\frac{ae}{r}\Big)^2\!\!\cos (2\xi)\nonumber \\  
&+&\frac{1}{8}\Big(\frac{ae}{r}\Big)^4\!\!\cos (4\xi) 
+\!\!\frac{5}{64}\Big(\frac{ae}{r}\Big)^6\cos (6\xi) 
+\frac{7}{128}\Big(\frac{ae}{r}\Big)^8\cos (8\xi)
+\cdots   \Big ].
\label{s1-27}
\end{eqnarray} 
The representation (\ref{s1-27})  with such ``separation of variables'' $(r,\xi)$ is useful
in an analytic calculation,  including a differentiation and an integration.
The number of terms, $\uk_{\rm max}$, in Eqs. (\ref{s1-24}),  (\ref{s1-27}) to be taken into account 
depends on the required precision.

The approximate formula (\ref{s1-27}) can now be  contrasted with an exact expression
of the electric field for a uniformly charged  elliptical beam. There is a compact formula 
\cite{Furman:1994, Furman:2007} in the complex $(x,y)$ plane, $z=x+  iy$, in the term of the ``complex electric field'':  
\beq
 E(z)\equiv E_x(z) +  i E_y(z) 
 = \frac{4\kappa q\lambda}{a^2-b^2}
(\bar{z}-\sqrt{\bar{z}^2-a^2+b^2}).
\label{s1-28}
\eeq
However, in the real components the formula is more complicated.
Outside the beam, the $x$-component of the field is
\beq
 E_x=\frac{4\kappa q\lambda}{(ae)^2}
\Big \{ x - \frac{sign(x)}{\sqrt{2}}\Big [u+\sqrt{u^2+(2xy)^2}\Big ]^{1/2}\Big \},
\label{s1-29}
\eeq
while the $y$-component can be obtained from this by exchanging $x\leftrightarrow  y$ and 
$a\leftrightarrow  b$. Here $u=x^2 - y^2 -(ae)^2$. 
Thus, 
\beq
 E_{\perp}(r,\xi)=\sqrt{E_x^2(r,\xi) + E_y^2(r,\xi)}
\label{s1-30}
\eeq
with $x=r\cos \xi$ and $y=r\sin \xi$.

We are now in a position to evaluate the number of terms  in Eq. (\ref{s1-27}) that
need to be taken into account in order to get the precision, say, to better than 5$\%$, 
if compared with the exact formula  (\ref{s1-30}). 
 Figure~\ref{fig2}(a) shows the variation of  the ratio of  Eq. (\ref{s1-27})  to  Eq. (\ref{s1-30})  with the azimuthal angle $\xi$ in the first quadrant.  
We observe that at $r/ae=1$ (full  curve) the desired precision is almost reached 
at $\uk_{\rm max}=7$,  except for the area near $\xi=0$. In this area, it is necessary to account for terms with $\uk_{max}>7$ to achieve  the required precision. 
At the same time, at $r/ae=2$ and $\uk_{max}=7$ (dashed  curve), the accuracy is better than 1$\%$ 
in the entire area of $\xi$.

The azimuthal field variation is essential only at $r \sim a$. 
For instance, for a flat beam and at $r = a$, the field is concentrated  at $\xi = 0$ and $\pi$. 
However, at larger $r$ (say, $r > 5a$) the angular dependence vanishes rapidly and the electric field of an ultra-relativistic elliptical bunch  converges to the universal form (\ref{s1-19}). 
The last statement is illustrated by Fig.~\ref{fig2}(b).


\section{Universality of a bunch external fields  }

The uniform particle density considered in the last section is an idealization. In reality, 
the linear particle density, 
$\lambda(z)$, varies considerably along the bunch. As an example,  Fig.~3(a) shows the current profile of 
electron bunches in the XFEL accelerator \cite{Altarelli:2006}. Electrons of energy 
17.5 GeV form bunches with a charge of 1 nC and a peak current of 5 kA. The current distribution is well 
fitted by a sum of two Gaussian distributions and a polynomial pedestal. Certainly, these bunches are 
ultra-relativistic since $\gamma = 3.2\times 10^4$. Now the question arises how to calculate the electric field 
of the bunch for a given distribution of the current density $J(z) = q\beta c\lambda(z)$.

\begin{figure*}[t]
  \centering
\begin{minipage}{.46\textwidth}
\includegraphics[width=\textwidth]{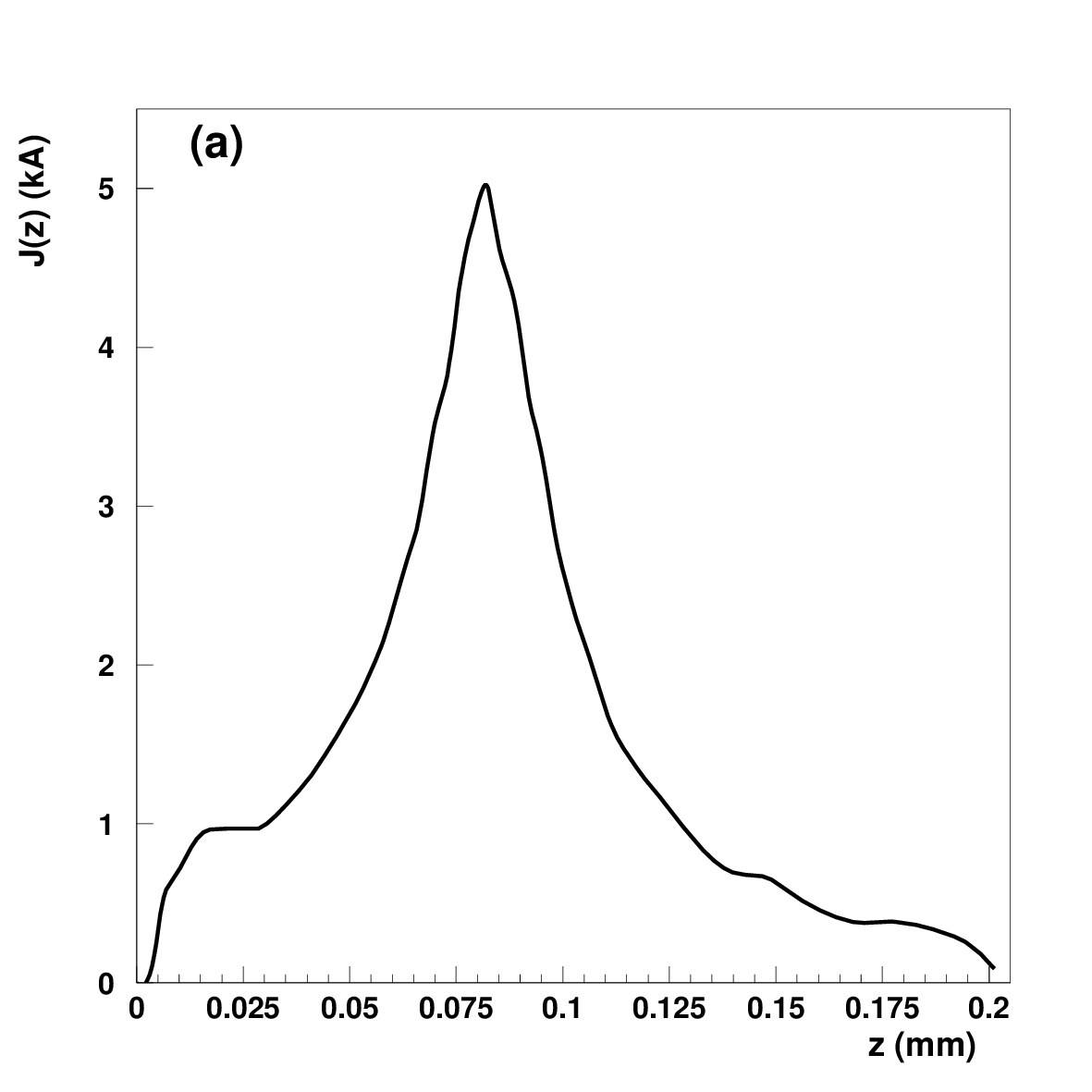}
\end{minipage}	\hspace*{+5mm}
\begin{minipage}{.46\textwidth}
\includegraphics[width=\textwidth]{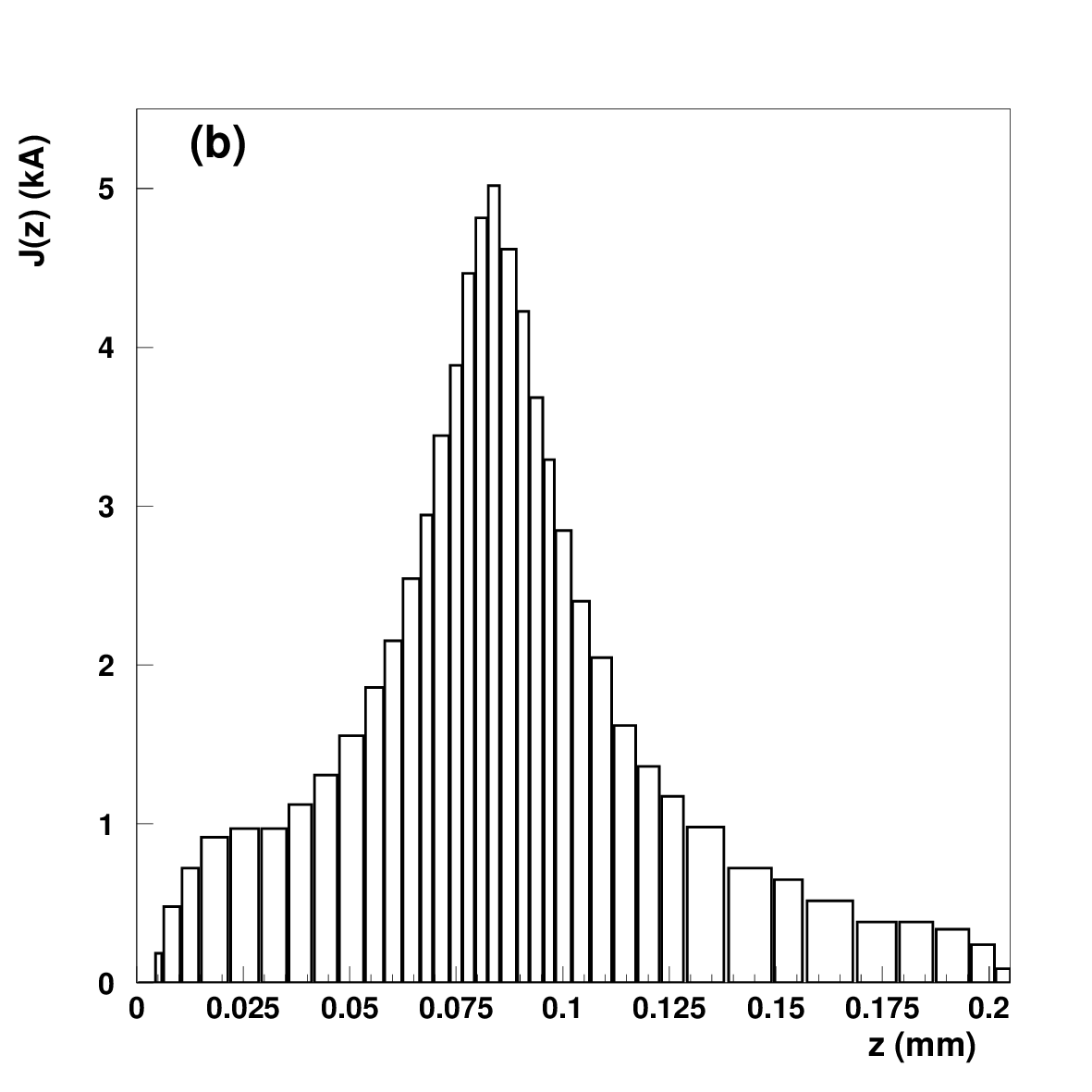}
\end{minipage}
\caption{ (a) The current profile of an electron bunch in the XFEL accelerator \cite{Altarelli:2006} and (b) its representation as a histogram with a variable bin size $\Delta z_i$.}
\label{fig3} 
\end{figure*}

We will now argue that in the asymptotic limit $\gamma \rightarrow \infty $ the problem has a simple solution.

The main conclusion that we can draw from the previous consideration is that at a distance of several bunch radii, the electric field of the bunch is independent of the transverse geometry of the bunch. 
Without loss of generality let consider again a bunch with a circular cross section. 
The distribution $J(z)$ can be well approximated by a histogram $J_\ui = q\beta c \lambda_\ui$, as shown in Fig.~\ref{fig3}(b). 
Let us imagine that the bunch is a set of layers with a thickness of $\Delta z_\ui$. 
Suppose that in the transverse direction particles are distributed uniformly and the linear particle density varies from one layer to the next in accordance with $\lambda_i$. 
Thus, each layer is a cylinder with a uniform particle density acting as an independent field source. 
The complete field of the bunch is a sum of elementary field sources (\ref{s1-19}) 
\beq
 E(r,z) = \kappa\frac{2q}{r}\sum_{\ui} \Big [\theta(z-z_\ui)-\theta(z-z_\ui- 
\Delta z_\ui)  \Big ]\lambda_\ui 
 =  \kappa\frac{2q}{r}\lambda (z,t).
\label{s3-1}
\eeq
Here we have used
\begin{displaymath}
 \lim_{\Delta z_\ui \to 0}\frac{\theta(z-z_\ui)-\theta(z-z_\ui- \Delta z_\ui)}{\Delta z_\ui}\Delta z_\ui=\delta(z-z_\ui)dz_\ui .
\nonumber 
\label{s3-2}
\end{displaymath}
and the sum is replaced by an integration. We obtain the same result (\ref{s3-1}) even if consider a bunch with a variable cross section and a uniform transverse density in each slice $\Delta z_\ui$.

We conclude with a statement that summarizes the obtained results:\\
{\bf Theorem}: 
{\it In the ultra-relativistic limit, $\gamma \rightarrow \infty $, the external electric 
field of a bunch with a linear particle density $\lambda (z)$ is governed by the universal law }
\beq
E(r,z,t)=\kappa\frac{2q}{r}\lambda (z,t).
\label{s3-3}
\eeq 

It is instructive to compare the strength of the electric field created by a cylindrical bunch with a uniform particle density $\lambda_U = N/L$ and a bunch with the Gaussian particle distribution $\lambda_G(z)$. The ratio of the fields is
\beq
\frac{E_G}{E_U}=\frac{\lambda_G(z)}{\lambda_U}=\frac{L}{L_{\rm eff}}\exp\Big(-\frac{z^2}{2\sigma^2_z}\Big).
\label{s3-4}
\eeq 
Thus, if $L$ is equal to the effective length of the Gaussian bunch $L_{\rm eff} = \sqrt{2\pi}\sigma_z$, the field strength in the both cases are equal at the maximum of $\lambda_G(z) $. 
Note, however, that in a more general case, as in Fig.~3, that conclusion is not correct, even if 
$L = L_{\rm eff}$. For instance, with parameters of a XFEL bunch \cite{Altarelli:2006}, one find  
$L_{\rm eff} = 0.217\,$mm and at the maximum of the current density $E_{\rm XFEL}/E_U$ = 3.63.

So far, we have considered fields in free space. In an accelerator, the charged beam is influenced by an environment and a high-intensity bunch induces surface charges or currents into this environment. 
This modifies the electric and magnetic fields around the bunch. 
There is a relatively simple method to account for the effect of the environment by introducing image charges and currents.


\section{\label{imch}Fields from image charges}

``{\it Definition of an electrical image.} An electrical image is
an electrified point or system of points on one side of a surface
which would produce on the other side of that surface the same
electrical action which the actual electrification of that surface
really does produce''\cite{Maxwell:1873a}.

Following Laslett \cite{Laslett:1963jn, Laslett:1987jn}, we consider 
a relativistic bunch between infinitely wide  conducting plates placed 
at $y= h$ and $y= -h$. 
Suppose that  the constituents of the bunch are positively charged. 
For full generality, let the  bunch be displaced by $(0,\bar{y},0)$ 
from the midplane $(x,0,z)$, and the observation point of the field be  at $(0,y,0)$
between  conducting parallel plates. 
The boundary condition for the  electric field  on perfectly conducting plates 
is $E_z(\pm h)=0$ 
and is satisfied if the image charges change sign from image to image. 

\begin{figure*}[t]
\includegraphics[height=10.928cm,width=12.cm]{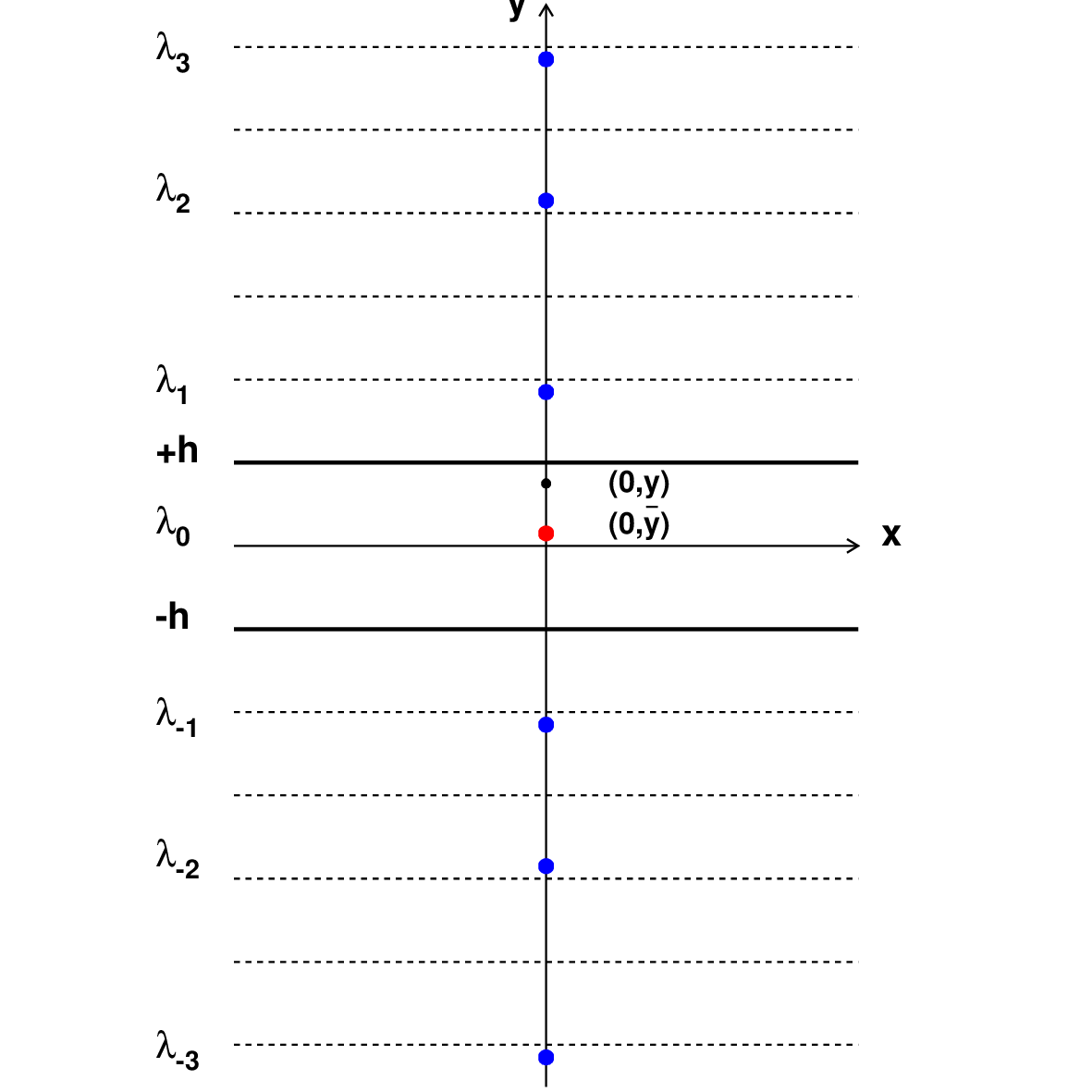}
\caption{
The electric field seen by a particle at location $y$ on the $y$-axis is generated by
the direct source-charge $\lambda_0 $ at  $\bar{y}$ and
the successive image charges $\lambda_{\pm \ui}$  at  locations   $d_{\pm \uk}$ and $d_{\pm \um}$ (see explanation in the text).  }
\label{fig4} 
\end{figure*}

The electric field seen by a particle at location $y$ on the $y$-axis is generated by
the direct source-charge $\lambda_0 $ and
the successive image charges $\lambda_{\pm \ui}$    \cite{Chao:1993zn, Hofmann:1992ki}, 
as shown in Fig.~\ref{fig4}. 
For instance, the image charges   $\lambda_1$ and $\lambda_{-1}$ are generated by $\lambda_0 $ due to 
reflection in  plates $+h$ and   $-h$, respectively. The image charges   $\lambda_2$ and $\lambda_{-2}$ 
are generated by $\lambda_{-1}$ and $\lambda_1$ due to reflection in  plates $+h$ and   $-h$, respectively,
and  so on. With the help of Fig.~\ref{fig4}, one can easily calculate the distance between the image charge 
position  and the observation point. 
So, for odd images, $\uk=1,3,5,...$, the distances  between $\lambda_{\pm \uk}$ and 
the point $y$ are  $d_{\pm \uk} = 2\uk h \mp y_1$. 
For even images, $\um=2,4,6,...$, the distances  
between $\lambda_{\pm \um}$ and the point $y$ are  $d_{\pm \um} = 2\um h \mp y_2$.
 Here $y_1=y+\bar{y}$ 
and $y_2=y-\bar{y}$. 

Suppose that the distance between plates is of the order $10a$.
Thus, the electric field of each image is described by Eq. (\ref{s3-3}).
To calculate the image electric field component $E_{y, \rm im} (y)$ in front of the plate, 
we add the contributions
from all image  fields in the infinite series \cite{Laslett:1963jn, Laslett:1987jn, Wiedemann:2007yf, Hofmann:1992ki}:
\begin{eqnarray}
E_{y, \rm im}(y,\bar{y},z,t)&=&2\kappa q\lambda (z,t)
[(2h-y_1)^{-1}\ -(2h+y_1)^{-1} 
- (4h-y_2)^{-1}\ +(4h+y_2)^{-1} \nonumber\\
&+&  (6h-y_1)^{-1}\ -(6h+y_1)^{-1}  
- (8h-y_2)^{-1}\ +(8h+y_2)^{-1} \nonumber\\
&+&  (10h-y_1)^{-1}-(10h+y_1)^{-1}
- (12h-y_2)^{-1}+(12h+y_2)^{-1} + ...],
\label{s4-1}
\end{eqnarray}
The representation (\ref{s4-1}) keeps  the same form irrespective of the
relative position of the bunch center and the observation point between plates,
$(\bar{y}\ge 0,\, y\ge\bar{y},\, y<\bar{y})$ or 
$(\bar{y}<0,\, y\le\bar{y},\, y>\bar{y})$.
These image fields  must be added to the direct 
field of the bunch (\ref{s3-3}) to meet the boundary  condition
that the electric field enters conducting surfaces perpendicularly.

In the original paper \cite{Laslett:1963jn} (see also Refs. \cite{Wiedemann:2007yf, Hofmann:1992ki}), the series (\ref{s4-1}) was summed up only in the linear
approximation in $y$ and $\bar{y}$:
\beq
E_{y,\rm im}(y,\bar{y})\,=\, \kappa\frac{4q\lambda}{h}\frac{\epsilon_1}{h}(y+2\bar{y}).
\label{s4-2}
\eeq
The coefficient $\epsilon_1=\pi^2/48$ is known as the Laslett coefficient (or form factor)
for infinite parallel plate vacuum chambers.
The approximation (\ref{s4-2})  widespread in textbooks and lectures,  
is, however, incorrect if the deviation of the bunch center
from the axis is large ($\bar{y}\sim h$) or if the field observation point  $y$ 
is located far off  the bunch.  
 Therefore, below we present the exact solution of the problem.

In  Appendix \ref{appA} it is proven that the exact summation of the series (\ref{s4-1}) gives
\beq
E_{y,\rm im}(y,\bar{y},z,t)\,=\,\kappa\frac{4q\lambda (z,t)}{h}\Lambda(\delta,\bar{\delta}),
\label{s4-3}
\eeq
where the electric image field structure function $\Lambda$  depends only on scaled variables
$\delta = y/h$,  $\bar{\delta} = \bar{y}/h$ in the form
\beq\label{s4-4} 
\Lambda (\delta,\bar{\delta})=\frac{1}{2} \Big [\frac{\pi}{2}\cdot \frac{\cos(\frac{\pi}{2}\bar{\delta})}
{\sin(\frac{\pi}{2}\delta)-\sin(\frac{\pi}{2}\bar{\delta})} -\frac{1}{\delta -\bar{\delta}}\Big ].
\eeq
We have shown in  Appendix \ref{appA}  that  the truncated linear approximation  
 (\ref{s4-3}) recovers the part (\ref{s4-2}) derived by Laslett.

We shall now calculate  values of   $\Lambda(\delta,\bar{\delta})$  at several particular 
points along the $y$-axis.
 
$\delta = 1$: the observation point is located at the plate, $y =h$. 
In this case, the structure function depends only on the bunch center position 
between plates, $\bar{\delta}$. From (\ref{s4-4}) one gets
\beq
\Lambda(1,\bar{\delta})= \frac{1}{2}\Big [\frac{\pi}{2} 
\frac{1+\sin(\frac{\pi}{2}\bar{\delta})}{\cos(\frac{\pi}{2}\bar{\delta})}
 - \frac{1}{1-\bar{\delta}}\Big ].
\label{s4-6} 
\eeq 
Equation (\ref{s4-6}) is singular at $\bar{\delta}\rightarrow 1$  and  shows that 
the conducting plate attracts the bunch with a force increasing
 with the bunch displacement from the midplane. The phenomenon, involving the transverse movement 
 of the bunch as a whole,  arises from image forces and could lead to a transverse instability of the beam.
This is discussed further in Sect. 6.

$\bar{\delta}=0, \delta = 1$: the bunch is in the midplane and the observation 
point is at the plate $y =h$
\beq 
\Lambda(1,0)= \frac{1}{2}(\frac{\pi}{2}-1).
\label{s4-7} 
\eeq 

\begin{figure*}
\begin{minipage}{.40\textwidth}
\includegraphics[width=\textwidth]{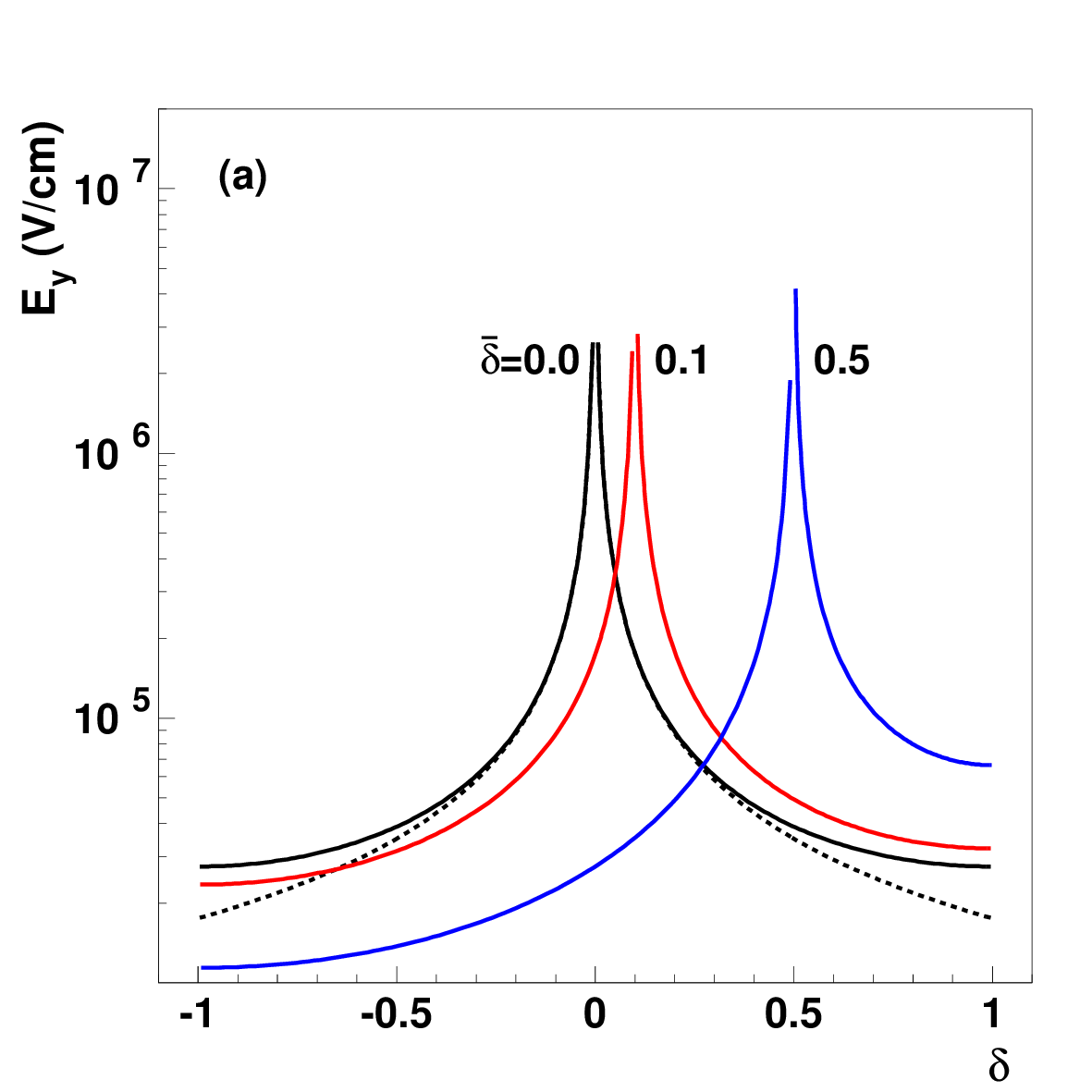}
\end{minipage}	\hspace*{+5mm}
\begin{minipage}{.40\textwidth}
\includegraphics[width=\textwidth]{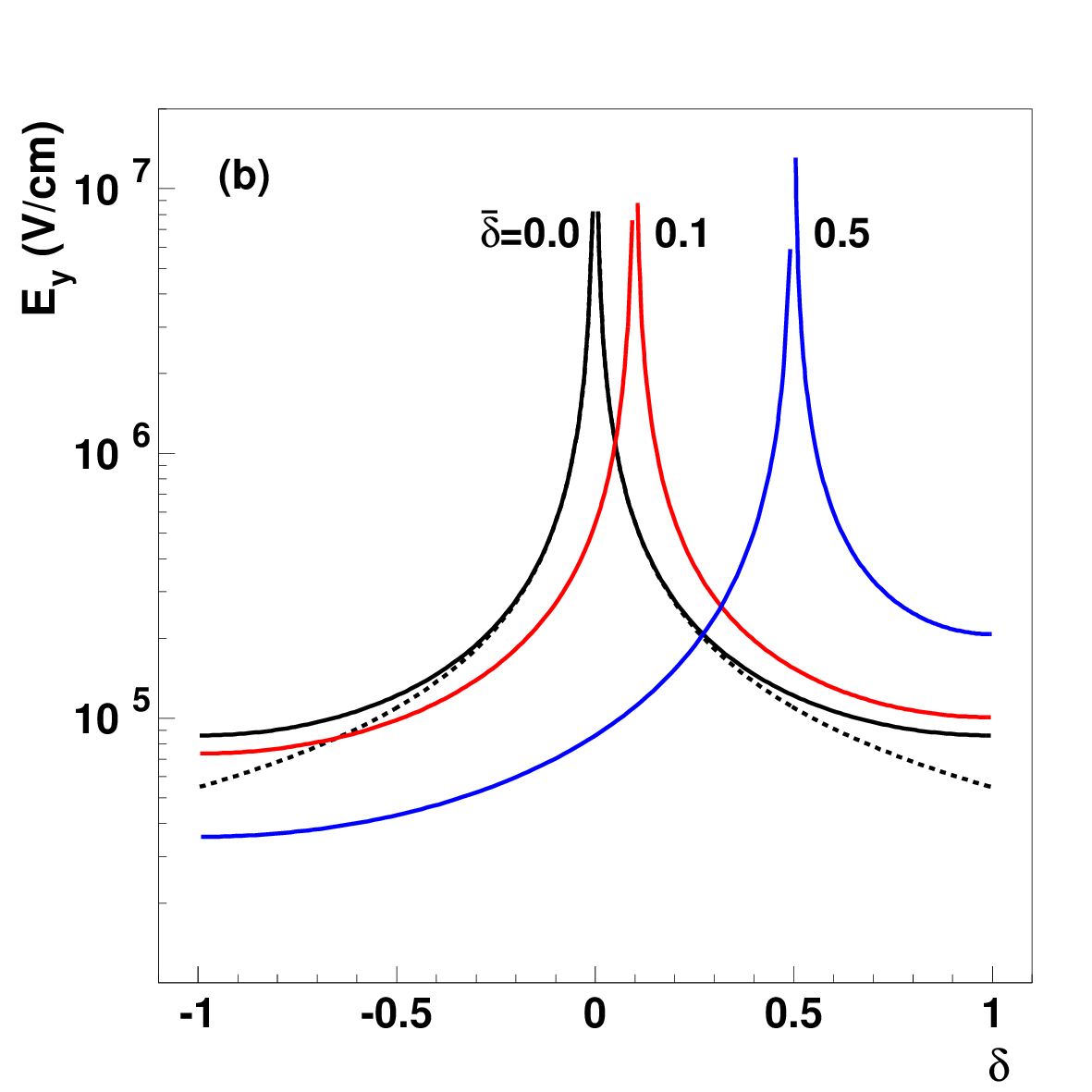}
\end{minipage}
\caption{ The electric field strength distribution
 in a gap between parallel conducting plates (solid curves) 
and in free space (dashed curves) at several values  of the bunch offset 
$\bar{\delta}$. 
(a) The LHC nominal proton beam scenario with parameters  
as they are given in Table 1. 
(b) An ILC-like positron beam, $N=2.0 \times 10^{10}$, $a = 17\mu$m, 
$\sigma_z$ = 300$\mu$m, and $h$ = 1.4 cm.}
\label{fig5} 
\end{figure*}

The image field (\ref{s4-3}) must be added to the direct 
field of the bunch (\ref{s3-3}) to meet the boundary  condition at conducting surfaces. 
It is interesting to note that the last term in Eq. (\ref{s4-4}) is opposite in sign 
to the direct field contribution outside the bunch and cancels it. As a result, 
the electric field distribution between parallel conducting plates is given by
\beq
E_{y,\rm tot}(y,\bar{y},z,t)= E_{y,\rm dir}+ E_{y,\rm im} 
=
\kappa\frac{q\lambda(z,t)}{h} 
\frac{\pi \cdot \cos (\frac{\pi}{2}\bar{\delta})}{\sin (\frac{\pi}{2}\delta)-
\sin (\frac{\pi}{2}\bar{\delta}) }.
\label{s4-8}
\eeq
In particular, a bunch moving in the midplane generates the field described by
\beq
E_{y,\rm tot}(y, 0,z,t)\,=\,\kappa\frac{2q\lambda(z,t)}{h} \cdot\frac{\pi/2}{\sin(\frac{\pi}{2}\delta )}.
\eeq
In other words, in the presence of conducting plates the electric field 
in front of the plate is enhanced by the factor $\pi/2$ (Fig.~\ref{fig5}).

If now we do not assume that the bunch offset $\bar{\delta}$  is small then the full linear approximation in $\delta$ can be derived by means of Eqs. (\ref{A1-8}) and (\ref{A1-14}) from  Appendix \ref{appA}. 
Thus, the vertical component of the electric  field seen by a test particle in the vicinity of the bunch ($|\delta-\bar{\delta}| \ll1$ ) is given by
\beq
E_{y,\rm tot}(y, \bar{y},z,t) \approx  \kappa\frac{2q\lambda(z,t)}{h}\Big 
[\frac{1}{\delta - \bar{\delta}}
+  \frac{\pi}{4}\tan (\frac{\pi}{2}\bar{\delta}) + 2\epsilon_1(\bar{\delta})(\delta-\bar{\delta})  \Big ]. 
\label{s4-9}
\eeq
Here we have introduced a generalization of the Laslett electric image coefficient $\epsilon_1$  in the case of an arbitrary offset:
\beq
\epsilon_1(\bar{\delta})\,=\,\frac{\pi^2}{32}\Big [\frac{1}{\cos^2(\frac{\pi}{2}\bar{\delta} )} 
- \frac{1}{3} \Big ],\qquad \epsilon_1(0)\,=\,\frac{\pi^2}{48}.
\label{s4-10}
\eeq
This approximation has to be compared with an alternative representation of Eq.
(\ref{s4-8})  in the form  (\ref{A1-8})
\beq
E_{y,\rm tot}(y, \bar{y},z,t) = \kappa\frac{2q\lambda(z,t)}{h}\frac{\pi}{4}
\Big \{ \tan \Big[\frac{\pi}{4}(\delta + \bar{\delta})\Big] 
+ \cot \Big[ \frac{\pi}{4}(\delta - \bar{\delta})\Big ] \Big \},
\label{s4-11}
\eeq
to show the origin of each term in Eq. (\ref{s4-9}). The potential function of the field 
(\ref{s4-11}) is
\beq
\mathrm{U}_{\rm tot}(y, \bar{y})= 2q\kappa\lambda\Big \{\ln \cos \Big[\frac{\pi}{4}(\delta + \bar{\delta})\Big]  
- \ln\sin \Big[ \frac{\pi}{4}(\delta - \bar{\delta})\Big ] \Big \}, 
\label{s4-12}
\eeq
with $E_{y,\rm tot}=-\partial\mathrm{U} /\partial y $.

In the linear approximation  one can obtain the horizontal component of the electric image field  directly from
\beq
\nabla \vec{E}_{\rm im} = \frac{\partial E_{x,\rm im}}{\partial x} 
+ \frac{\partial E_{y,\rm im}}{\partial y} = 0,
\label{s4-4a} 
\eeq 
with the use of Eqs. (\ref{A1-14}) and (\ref{s4-3}). Thus,
\beq
E_{x,\rm  im}(x, \bar{y},z,t) \approx - \kappa\frac{4q\lambda(z,t)}{h}\epsilon_1(\bar{\delta}_y)\delta_x,
\label{s4-13}
\eeq 
\beq
E_{y,\rm im}(y, \bar{y},z,t) \approx  \kappa\frac{4q\lambda(z,t)}{h}\Big
[ \frac{\pi}{8}\tan (\frac{\pi}{2}\bar{\delta}_y) + \epsilon_1(\bar{\delta}_y)(\delta_y-\bar{\delta}_y)  \Big ],
\label{s4-14}
\eeq 
with $\delta_x=x/h$, $\delta_y=y/h$ and $\bar{\delta}_y=\bar{y}/h$.
As follows from (\ref{s4-2}) and (\ref{s4-3}), the image fields produce defocusing forces in the $y$-direction.
On the other hand, due to (\ref{s4-13}), the corresponding forces in the $x$-direction produce focusing forces.

Equations (\ref{s4-9}) and (\ref{s4-11}) tell us that with an increase of $\bar{\delta}_y$, 
the field strength near the bunch and the field gradient across the bunch, 
$\partial E_{y,\rm im}/\partial y \sim 1/\cos^2(\frac{\pi}{2}\bar{\delta}_y )$, significantly increase.  
This is illustrated by Fig.~\ref{fig5}, which  shows that  
 with an increase of $\bar{\delta}_y$ the field distribution
between plates becomes more and more asymmetric. 
At the opposite ends of the bunch diameter
the difference in the value of the field, $\Delta E_{y,\rm tot}(\bar{\delta}_y)$, 
grows as the displacement increases.
For the LHC beam one get $\Delta E_{y,\rm tot}(0.1)$= 4360 V/cm,  
$\Delta E_{y,tot}(0.5)$ = 27290 V/cm, and for
the ILC beam  $\Delta E_{y,\rm tot}(0.1)$= 13640 V/cm and  $\Delta E_{y,\rm tot}(0.5)$= 85340 V/cm, 
respectively. In Sect. 6 we discuss how this effect modifies the tune shifts.


\section{\label{imcu}Magnetic images}

In the above, we have discussed electric image fields created by an ultra-relativistic bunch. Magnetic images can be treated in much the same way \cite{Wiedemann:2007yf}, \cite{Hofmann:1992ki}. 
Let the ferromagnetic boundaries be represented by a pair of infinitely wide
parallel plates at $y= +g$ and $y= -g$. 
The magnetic field lines must enter the magnet pole faces perpendicularly.
For magnetic image fields we distinguish  between DC and AC image fields.
The DC field penetrates the metallic vacuum chamber and reaches the ferromagnetic poles.
In case of bunched beams the AC fields are of rather high frequency, and we assume that they do not penetrate the thick metallic vacuum chamber. 
The $DC$ Fourier component of a bunched beam current is equal to twice 
the average beam current $\uJ =qc\beta\lambda \Bc$ \cite{Wiedemann:2007yf}, 
where $\Bc$ is  the Laslett bunching factor. 

A magnetic field, seen by a particle at location $y$ on the $y$-axis, is generated by the
successive image currents with the same sign as the beam itself. In Appendix \ref{appB} it is proven
that the resulting field is described by
\beq
B_{x,\rm im, DC}(y,\bar{y},z)\,=\,\frac{4\kappa q\beta \lambda(z)}{gc}\Bc\cdot H(\eta_y,\bar{\eta}_y). 
\label{s5-1}
\eeq
Here we have made the replacement $\mu_0=1/(\epsilon_0 c^2)$ and used the scaled variables $\eta_y=y/g $
and $\bar{\eta}_y=\bar{y}/g $, $\Bc=n_bL/2\pi R$ is the bunching factor, $n_b$ the number of bunches, $R$ the average accelerator radius. The structure function $H$ is of the form
\beq
H(\eta_y,\bar{\eta}_y)\,=\,\frac{1}{2}\Big [\frac{1}{\eta_y -\bar{\eta}_y}- \frac{\pi}{2}\cdot
\frac{\cos(\frac{\pi}{2}\eta_y)}{\sin(\frac{\pi}{2}\eta_y)-\sin(\frac{\pi}{2}\bar{\eta_y})} \Big ].
\label{s5-2}
\eeq
In the functional sense, $H(\eta_y,\bar{\eta}_y)\,=\,\Lambda (\bar{\eta}_y,\eta_y )$,
as can be noticed by comparing Eqs. (\ref{s5-2}) and (\ref{s4-4}).

In the linear approximation in $y$ and $\bar{y}$ one obtains from the exact formula (\ref{s5-2})
(see Appendix \ref{appB} for details)
\beq
B_{x,\rm im,DC}(y,\bar{y})\,\simeq\,\frac{4\kappa q\beta \lambda(z)}{g^2c}\Bc\epsilon_2\cdot (y+\frac{1}{2}\bar{y}),
\label{s5-3}
\eeq
where $\epsilon_2=\pi^2/24$ is the Laslett form factor for infinite
parallel plate magnet 
poles\footnote{In Ref. \cite{Wiedemann:2007yf} Eq. (18.57) should be read with the factor $(y + \bar{y}/2)$ as in (\ref{s5-3}). }. 
 As above for electric
images, we define a generalized form of $\epsilon_2$ for an arbitrary offset $\bar{\eta}_y$ as follows
\beq 
\epsilon_2 (\bar{\eta}_y)\,=\,\frac{\pi^2}{32}\Big [\frac{1}{\cos^2(\frac{\pi}{2}\bar{\eta}_y)}
 + \frac{1}{3}  \Big ],\qquad \epsilon_2 (0)\,=\,\frac{\pi^2}{24}.
\label{s5-4}
\eeq
Thus, the complete linear approximations in $\eta_x$ and $\eta_y$ (see Appendix \ref{appB})
are given by
\beq
B_{x,\rm im,DC}(y,\bar{y},z)\,\simeq  \,\frac{4\kappa q\beta \lambda(z)}{gc}\Bc
\Big [ \frac{\pi}{8}\tan ( \frac{\pi}{2}\bar{\eta}_y)
+\epsilon_2 (\bar{\eta}_y)(\eta_y - \bar{\eta}_y) \Big ]
\label{s5-5}
\eeq
and
\beq
B_{y,\rm im,DC}(x,\bar{y},z)\,\simeq \,\frac{4\kappa q\beta \lambda(z)}{gc}\Bc
\epsilon_2 (\bar{\eta}_y)\eta_x.
\label{s5-5a}
\eeq

For further applications, we point out that on the bunch axis, $\eta_y =\bar{\eta}_y$,   from Eq. (\ref{C-5}) one gets
\beq
H(\bar{\eta}_y,\bar{\eta}_y)\,=\,\frac{\pi}{8}\tan(\frac{\pi}{2}\bar{\eta}_y).
\label{s5-6}
\eeq

The contribution of the magnetic AC image field due to eddy currents in vacuum 
chamber walls is similar to electric image fields,
\beq
B_{x,\rm im,AC}(y,\bar{y},z)\,=\,- \frac{4\kappa q\beta \lambda(z)}{hc}(1-\Bc)\cdot
\Lambda(\delta_y,\bar{\delta}_y),
\label{s5-7}
\eeq
and therefore
\beq
B_{y,\rm im,AC}(x,\bar{y},z)\,=\,- \frac{4\kappa q\beta \lambda(z)}{hc}
(1-\Bc)\epsilon_1(\bar{\delta}_y)\delta_x,
\eeq
where the factor $(1-\Bc)$ accounts for the subtraction of the DC component.
Thus, at $\delta_x=0$ the net AC field is tangential to the surface.

The magnetic image fields must be added to the direct magnetic field (\ref{I-2}) 
to meet the boundary condition at ferromagnetic surfaces. That is, 
the summary horizontal component of the magnetic field  between the conducting plates is
\begin{eqnarray}
&&B_{x,\rm tot}(y,\bar{y}) = B_{x,dir} +B_{x,\rm im,DC} + B_{x,\rm im,AC} \nonumber \\   
&=&  -\frac{\pi \kappa q\beta \lambda}{hc}\bigg \{
\frac{(1-\Bc)\cos[(\pi /2)\bar{\delta}_y]\theta (1-\delta_y) }
{\sin [(\pi /2)\delta_y]-\sin [(\pi /2)\bar{\delta}_y]}
+ \frac{h}{g}\cdot\frac{\Bc\cos[(\pi /2)\eta_y]}{\sin[(\pi /2)\eta_y]
-\sin [(\pi /2)\bar{\eta}_y]}
\bigg \}.
\label{s5-8}
\end{eqnarray}
The step function $\theta (1-\delta_y)$ accounts for the fact that the AC fields do not penetrate the thick metallic vacuum chamber. For a more detailed discussion of the subject see Refs. \cite{Zotter:1972, Petracca:1999}.


\section{Image forces and tune shifts}

Direct space-charge fields, as well as fields due to image charges and currents shift 
the betatron frequencies (tunes). 
We have to distinguish between  coherent tune shifts, 
which express a change of the betatron frequency when the bunch oscillates 
as a whole, and  incoherent tune shifts, which change the single particle tune. 
In this section we again assume the bunch to have a circular cross section of 
radius $a$ and a uniform density.

In the next two subsections we restrict our analysis to an idealized case of a
vacuum chamber and/or ferromagnetic poles as two infinite parallel plates and
the motion of a bunch in the vertical 
$y$-direction\footnote{For this reason, 
below we omit the subscripts $x$ and $y$ for variables $\delta$ and $\eta$.
 If there are no ferromagnetic poles, one has to set $g=\infty$ 
in  each  equation below.}.
Infinite parallel plates are a good approximation for finite-width collimator parallel plates, 
given the small transverse beam size in modern accelerators.
Our equations for tune shifts are valid at an arbitrary bunch offset 
 and presented in notations as given 
in the  textbook \cite{Chao:1993zn} and the handbook \cite{Zotter:1999b}.
This allows us to compare  results obtained by different authors.
In Sect. 6.3 we consider a more realistic example of a finite-length collimator with
parallel conducting jaws.
\subsection{Coherent motion and tune shift}
%

The motion of the bunch center $\bar{\delta}(s)$ in the absence of an external 
focusing force is described by the equation
\beq
\frac{\ud^2\bar{\delta}(s)}{\ud s^2}= \frac{F_{y,im}}{M_b\gamma h \beta^2 c^2},
\label{s6-1}
\eeq
where $s = \beta ct$ and the Lorentz force is of the form
\begin{eqnarray} 
F_{y,im}  &=&  Q_b(E_{y,im} +\beta c B_{x,im}) \nonumber \\ 
&=& \frac{4 \kappa Q_b q\lambda}{h}\Big [ \Big ( \frac{1}{\gamma^2} 
+ \beta^2\Bc\Big )\Lambda(\delta,\bar{\delta})
+ \beta^2(h/g)\Bc H(\eta,\bar{\eta}) \Big ].
\label{s6-2}
\end{eqnarray}
Here we have applied the results of the previous section; $M_b=Nm_p$ is the bunch mass,
$Q_b= Nq$ the bunch charge.
On the bunch axis, the electric and magnetic structure functions are
\beq
\Lambda(\bar{\delta},\bar{\delta})=\frac{\pi}{8}\tan(\frac{\pi}{2}\bar{\delta}),\qquad
H(\bar{\eta},\bar{\eta})= \frac{\pi}{8}\tan(\frac{\pi}{2}\bar{\eta}).
\label{s6-3}
\eeq
\begin{figure*}
\begin{minipage}{.45\textwidth}
\includegraphics[width=\textwidth]{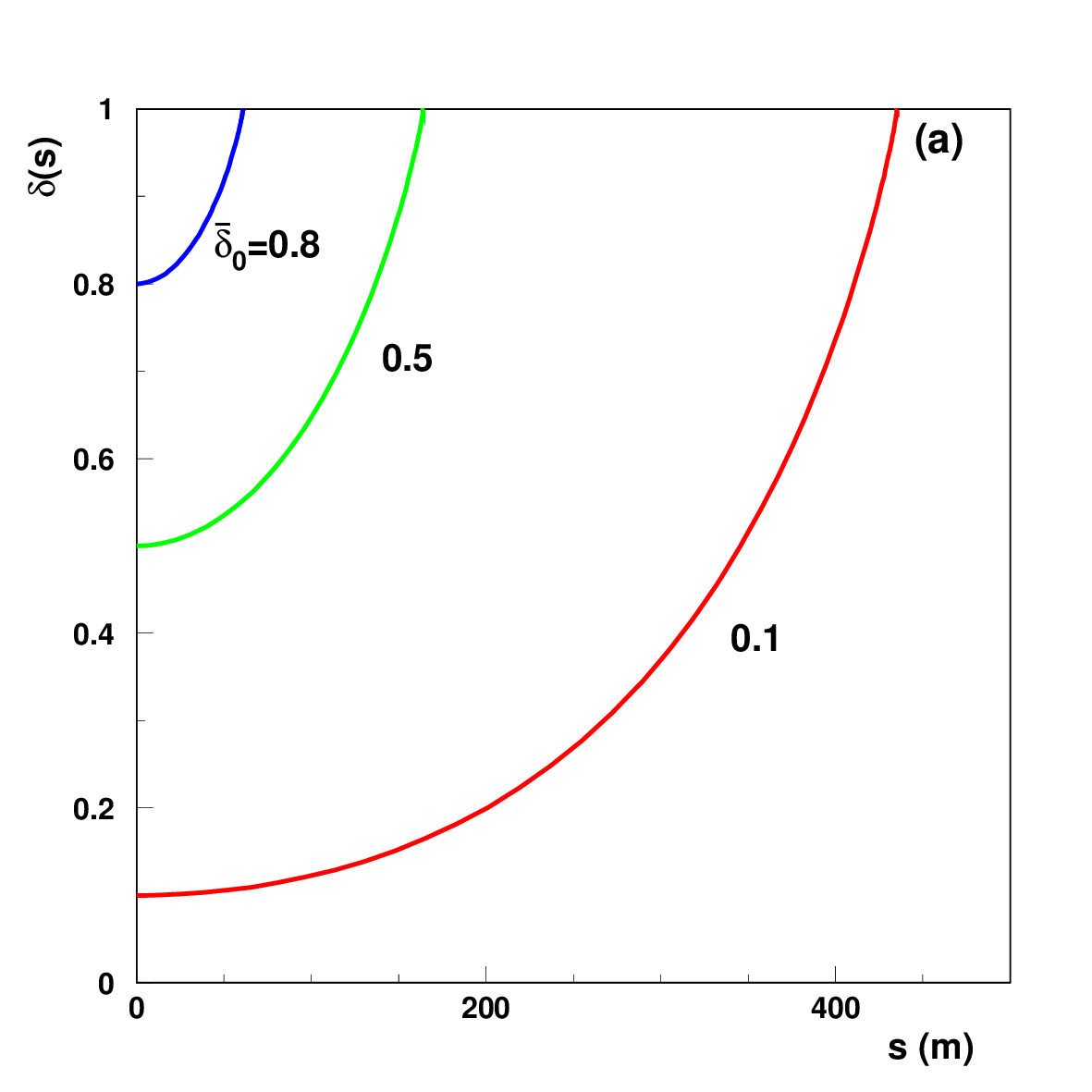}
\end{minipage}	\hspace*{+2mm}
\begin{minipage}{.45\textwidth}
\includegraphics[width=\textwidth]{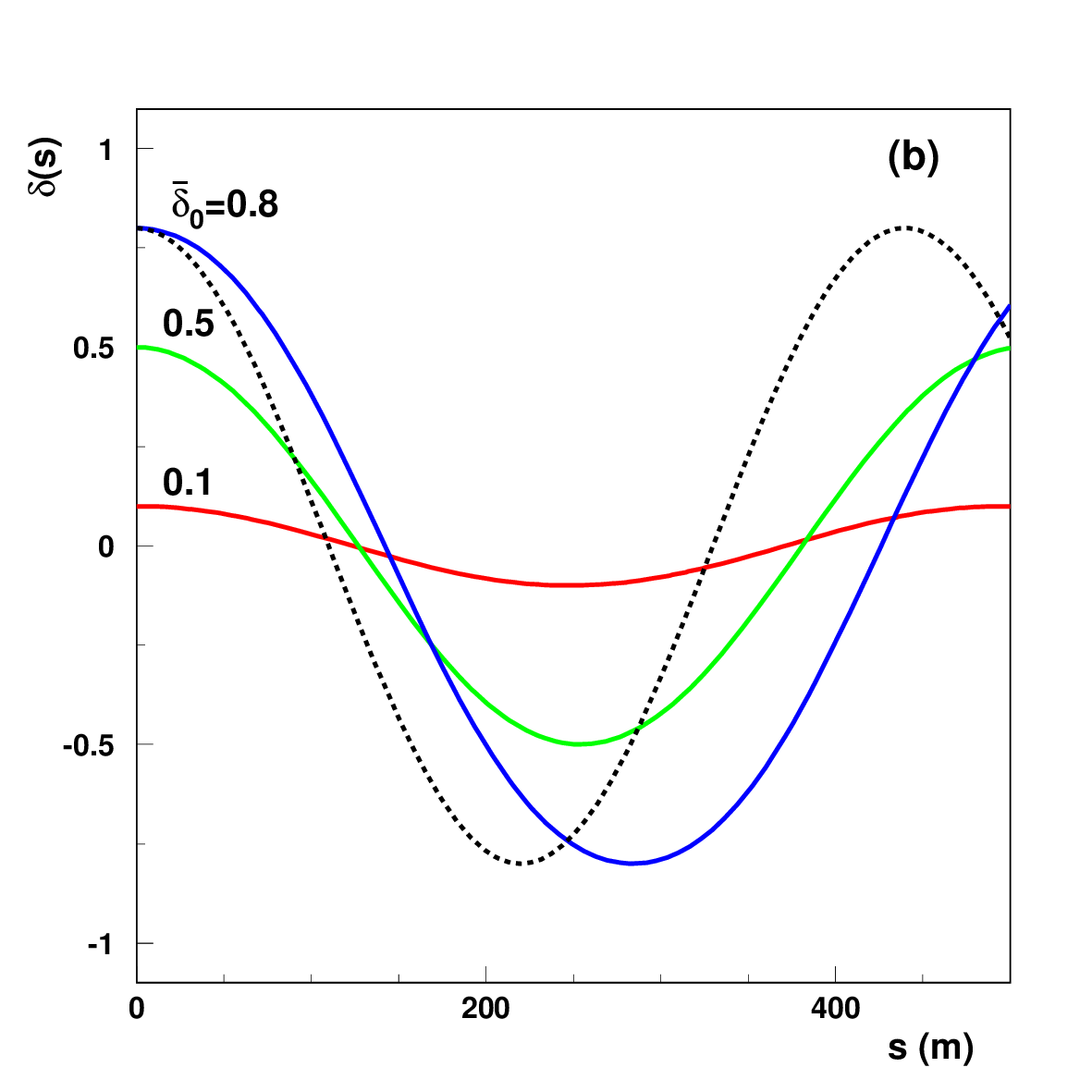}
\end{minipage}
\caption{ Numerical solutions of Eqs. (\ref{s6-1}) and (\ref{s6-4}). 
The dependence of the bunch trajectory along the beam path  on 
the initial value $\bar{\delta}_{0}$. 
(a) The linear focusing is switched off,
transverse motion only under the influence of the image forces.
The bunch impacts the plate at $\delta (s_\ui) =1$.
(b) The coherent oscillation of the bunch under the influence of the linear 
focusing and image forces (solid lines). 
The dashed line shows the betatron oscillation at $\bar{\delta}_{0}= 0.8$ 
without taking account of the image forces, $\Ic = 0$.}
\label{fig6} 
\end{figure*}

Under the influence of the force (\ref{s6-2}) the bunch is attracted by a conducting plate 
and at some point $s_i$ hits the plate. The actual position of the impact point 
depends on the initial value constraints, in particular, the bunch offset 
$\bar{\delta}(0)=\bar{\delta}_{0}$. Figure~\ref{fig6}(a) shows the numerical 
solutions of Eq. (\ref{s6-1}) with a set of initial conditions  
($\bar{\delta}_{0}$, $\bar{\delta}^{\prime}_{0} =0$) 
at  $\bar{\delta}_{0}= 0.1,\, 0.5,\, 0.8\,$. For example, let protons in the bunch be 
at the energy 7 TeV  and the machine parameters are as in Table 1 \cite{Bruning:2004}.
Then the impact points $s_\ui(\bar{\delta}_{0})$ are located at  distances
$s_\ui(0.1)=453.3\, m$,  $s_\ui(0.5)=164.2\, m$ and $s_\ui(0.8)=61.0\, m$, respectively. 

It should be noted that Eq. (\ref{s6-1}) is true in the approximation,
when the bunch velocity along the $z$-axis is constant, and much larger than the drift velocity  in the $y$ direction, $v_z \simeq \beta c\gg v_y$. In this case, the bunch path is a smooth curve, as shown in Fig.~\ref{fig6}(a). However, with a more rigorous account of the effect of crossed electric and magnetic fields from the images, it is necessary to solve together a system of two coupled differential equations for movements along the $z$ and $y$ directions. In this case, the bunch trajectory is a cycloid-like curve.

The coherent motion of the bunch in the $y$-direction is significantly altered in the presence of the linear
focusing provided by  quadrupoles and described by the equation
\beq
\frac{\ud^2\bar{\delta}(s)}{\ud s^2} + 
K^2_0 \bar{\delta}(s)
-\Ic \bigg \{ \frac{\pi}{2}\Big ( \frac{1}{\Bc \beta^2\gamma^2} +1 \Big )
\tan \Big [\frac{\pi}{2}\bar{\delta}(s)\Big ]
+ \frac{\pi h}{2g}\tan \Big [\frac{\pi h}{2g}\bar{\delta}(s)\Big ] \bigg \} = 0.
\label{s6-4}
\eeq
Here
\beq
K^2_0=\Big (\frac{\nu_{0}}{R} \Big )^2 , \qquad
\Ic= \frac{ r_p \lambda \Bc}{ h^2 \gamma}, \nonumber
\eeq
where $r_p= \kappa q^2/m_p c^2$ is the classical proton radius and 
the meaning of the other parameters is explained in Table 1. With the values of 
the parameters\footnote{With $\nu_{y0}\simeq \nu_{x0}=\nu_0$.}
 from Table 1, $K^2_0 = 2.041\times 10^{-4}$ m$^{-2}$ and
$\Ic =1.73\times 10^{-5}$ m$^{-2}$. Thus, for small $\bar{\delta}_{0}$ 
the linear focusing is a driving force.

Figure~\ref{fig6}(b) shows numerical solutions of Eq. (\ref{s6-4}) for the same 
initial conditions as above. The dashed line shows a solution of the betatron 
equation (\ref{s6-4}) in the absence of the image effects, $\Ic = 0$. 
A comparison of the two curves at $\bar{\delta}_{0}=0.8$ demonstrates how big  the
influence of images on the coherent tune shift is.

To derive an analytical expression for the coherent tune shift for an arbitrary 
offset we proceed in the standard way \cite{Chao:1993zn, Hofmann:1992ki, Zotter:1972}. 
In the linear theory, we assume that the forces are proportional to the displacement. 
Therefore, we expand the structure functions $\Lambda$ and $H$ (\ref{s6-3}) 
in a power series in the neighborhood of $\bar{\delta}_0$, 
$\bar{\delta}= \bar{\delta}_0 + \Delta$, keeping only terms up to first
order in $\Delta$:
\beq
 \Lambda(\bar{\delta}_0, \Delta) = \frac{\pi}{8}\tan \Big [\frac{\pi}{2}(\bar{\delta}_0+\Delta)\Big ]
 \approx  \frac{\pi}{8}\tan \Big (\frac{\pi}{2}\bar{\delta}_0\Big ) + \xi_1(\bar{\delta}_0)\Delta(s),
\label{s6-5}
\eeq
%
%
\beq 
H(\bar{\eta}_0,\Delta)\approx \frac{\pi}{8}\tan \Big (\frac{\pi}{2}\bar{\eta}_0 \Big )
+ \frac{h}{g}\xi_2(\bar{\eta}_0)\Delta(s) .
\label{s6-6}
\eeq
Here we have introduced generalized Laslett coherent tune shift form factors,
\begin{eqnarray}
 \xi_1(\bar{\delta})&=& \frac{\pi^2}{16\cos^2(\frac{\pi}{2}\bar{\delta})},\qquad
\xi_1(0)=\frac{\pi^2}{16},\nonumber \\
 \xi_2(\bar{\eta}) &=& \frac{\pi^2}{16\cos^2(\frac{\pi}{2}\bar{\eta})},\qquad
\xi_2(0)=\frac{\pi^2}{16},
\label{s6-6-1}
\end{eqnarray}
for the image fields from the vacuum chamber and the magnet pole. 
Substituting  expressions for $ \Lambda(\bar{\delta}_0, \Delta)$ and 
$H(\bar{\eta}_0,\Delta) $ in (\ref{s6-4}), 
we get\footnote{We add a hat to the amplitude function
$\hat{\beta}$ to avoid confusion with the relativistic velocity $\beta$.}
\beq
 \Delta\nu^{(\rm coh)}_y(\bar{\delta}_0)  = -\,
\frac{R \langle \hat{\beta} \rangle}{2m_pc^2\gamma\beta^2}\frac{\partial F_{y,\rm im}}{\partial y} 
  =  -\, \frac{2r_p \uJ R \langle \hat{\beta} \rangle }{q\beta c\gamma}
\Big [\Big (\frac{1}{\Bc\beta^2\gamma^2} +1\Big )\frac{\xi_1(\bar{\delta}_0)}{h^2}
+ \frac{\xi_2(\bar{\eta}_0)}{g^2} \Big ].
\label{s6-7}
\eeq

One must note that the image coefficients $\epsilon_1$ and $\xi_1$, as well as 
 $\epsilon_2$ and $\xi_2$,  are not independent but are rooted in the same function 
$\Lambda(\delta, \bar{\delta})$ and $H(\eta,\bar{\eta})$,  correspondingly.
Therefore, in the linear approximation these functions are related via
\beq 
\epsilon_1(\bar{\delta})=\frac{1}{2}\Big [ \xi_1(\bar{\delta}) - \frac{\pi^2}{48} \Big ],\qquad \epsilon_2(\bar{\eta})=\frac{1}{2}\Big [ \xi_2(\bar{\eta}) + \frac{\pi^2}{48} \Big ]\, .
\label{s6-8}
\eeq

\begin{table}[!h]
\renewcommand{\arraystretch}{1.3}
\begin{center}
\caption{The LHC machine and beam parameters \cite{Bruning:2004} 
used in  calculation of the coherent and incoherent tune shifts.}\label{LHC_par}
\vskip0.2in
\small  
\begin{tabular}{@{}|l||c||c|}
\hline
$h$, collimator half-gap &[m]  & $1.2\times 10^{-3}$\\\hline
$g$, magnet poles half-gap & [m] & $4.0\times 10^{-2}$\\\hline
$N$, bunch population &  &$1.15\times 10^{11}$\\\hline
$\sigma_z$, r.m.s. bunch length & [m] &$7.55\times 10^{-2}$ \\\hline
$a$, r.m.s bunch radius &[m]  &$1.67\times 10^{-5}$ \\\hline
$\Bc$, bunching factor &  &0.1993 \\\hline
$\lambda = N/\sqrt{2\pi}\sigma_z$, linear density &  & \\\hline
$J=q\beta c \lambda\Bc$, average beam current &  & \\\hline
$m_p$, proton mass &[GeV]  &0.938 \\\hline
$\gamma$, Lorentz factor &  & 7463\\\hline
$\nu_0=R/\langle \hat{\beta} \rangle $, betatron tune&  &60.61 \\\hline
$\langle \hat{\beta} \rangle$, average $\hat{\beta}$-function  &[m]  &70 \\\hline
$2\pi R$, ring circumference & [m] & 26\,658.883\\\hline
\end{tabular}
\end{center}
\end{table}

\subsection{Incoherent  tune shifts}

Let us now evaluate the effect of the image forces on the betatron oscillation of 
the particles in a bunch.
The motion of a test particle in a displaced bunch in the presence of 
the space-charge
force and the image fields is described by the equation
\beq
\frac{\ud^2\delta}{\ud s^2} + K^2_0 \delta
= \frac{(F_{y,\rm sc} + F_{y,\rm im})}{mh\gamma\beta^2c^2}.
\label{s6-9}
\eeq
Here $F_{y,\rm sc}=2\kappa q^2\lambda y/a^2\gamma^2 $ is the Lorentz force due 
to the bunch space-charge \cite{Wiedemann:2007yf} and $F_{y,\rm im} $ is defined 
in (\ref{s6-2}). 
Inserting the linear approximations (\ref{A1-14}) and (\ref{C-7}) 
into Eq. (\ref{s6-9}), in the same manner as above one gets a vertical tune shift
\beq
\Delta\nu^{(\rm inc)}_y(\bar{\delta})
\! = \! -\, \frac{2r_p \uJ R \langle \beta \rangle }{q\beta c\gamma}
\Big [\frac{1}{\Bc\beta^2\gamma^2}\Big (\frac{1}{2a^2} + 
\frac{\epsilon_1(\bar{\delta})}{h^2}\Big ) 
+ \frac{\epsilon_1(\bar{\delta})}{h^2}
+ \frac{\epsilon_2(\bar{\delta})}{g^2} \Big ].
\label{s6-10}
\eeq
The above analysis can be carried out similarly for the $x$-motion. The result is
\beq
\Delta\nu^{(\rm inc)}_x(\bar{\delta})
\! = \! \frac{2r_p \uJ R \langle \beta \rangle }{q\beta c\gamma}
\Big [\frac{1}{\Bc\beta^2\gamma^2}\Big (\frac{\epsilon_1(\bar{\delta})}{h^2} - 
\frac{1}{2a^2} \Big ) 
+ \frac{\epsilon_1(\bar{\delta})}{h^2}
+ \frac{\epsilon_2(\bar{\delta})}{g^2} \Big ].
\label{s6-11}
\eeq
Equations (\ref{s6-7}), (\ref{s6-10}), and (\ref{s6-11}) generalize the Laslett tune 
shifts to the case of the arbitrary bunch offset between parallel conducting plates
 and ferromagnetic poles.  

As numerical examples, with machine and beam parameters from Table 1, let us compare 
the contribution of each term in Eq. (\ref{s6-10}) at two
distinct values of $\bar{\delta}$, $\bar{\delta}=0$ and $\bar{\delta}=0.8$
\begin{displaymath}
\Delta\nu^{(\rm inc)}_y(0)
 = -\!\! (3.8\times 10^{-4}+1.9\times 10^{-7}
+ 2.113 + 3.8\times 10^{-3})\approx -2.117, 
\end{displaymath}
\begin{displaymath}
\Delta\nu^{(\rm inc)}_y(0.8)
\! = \!\!-\!\! (3.8\times 10^{-4}+2.9\times 10^{-6}
+ 32.136 + 3.81\times 10^{-3})\approx -32.14\,. 
\end{displaymath}
Thus, the third term gives the main contribution that 
increases with $\bar{\delta}$, and the 
transverse particle dynamics in a bunch is defined by the influence of electric images.

\subsection{Collimator of a finite length}

The representation of a vacuum chamber and magnetic poles in the form 
of infinite parallel plates is a very useful mathematical abstraction.
However, in real accelerators, all components are finite in size
and at the same time some of these components include elements that are structurally designed as
parallel conductive and ferromagnetic flat surfaces.
In circular accelerators such as LHC \cite{Bruning:2004}  and the future HL-LHC \cite{Bruning:2017}, 
flat parallel surfaces are  parts of different types of collimators, 
 the normal conducting  separator and orbit correction dipole 
magnets\footnote{A list of collimators for the LHC Run 2 (in 2015) includes  108 items
and shown on p. 151 of the technical design report 
``High-Luminosity Large Hadron Collider (HL-LHC)'' \cite{Bruning:2017}.}.
As a rule, collimator jaws have a length of 600--1400 mm, 
and their width is about 50 mm.
Similarly, the poles of a dipole magnet 
have a length of 2000--3400 mm and a pole  width of 60 mm \cite{Bruning:2017}.
With transverse beam sizes as small as 200 $\mu$m, the representation of
collimators and dipole magnets in the form of infinite  parallel 
plates is a good approximation for these elements
and it is legitimate to apply here the results obtained in the previous sections.

Suppose that the accelerator ring only includes one collimator at point $s_0$.
A collimator is a short straight section of an accelerator 
and does not have a guide magnetic field. The motion of the bunch
center $\bar{\delta}$ in the collimator is therefore described by Eq. (\ref{s6-1}),
with $F_{y,\rm im}$ from Eq. (\ref{s6-2}), where we set $g=\infty$. In the rest part of 
a circular accelerator the motion of the bunch is described 
by Eq. (\ref{s6-4}) with $\Ic=0$. 

Now we would like to
derive in the framework of linear theory an analytical expression
for the  coherent tune shift due to the image effects in a finite-size
collimator. For more than a single collimator one would simply add the individual
contribution from each collimator to find the total tune shift.

The  image fields $E_{y,\rm im}$ and $B_{x,\rm im,AC}$   act along the same line,
so we introduce the effective field $B_{eff}$ and decompose it into  dipole-like 
and  quadrupole-like parts with the use of Eqs. (\ref{s4-3}), (\ref{s5-7}) and (\ref{s6-5}),
\beq
B_{\rm eff} = B_{x,\rm im,AC} + E_{y,\rm im}/\beta c = B_D + G_{\rm eff}\cdot \Delta (s),
\label{s63-1}
\eeq
where
\beq
B_D = \kappa\frac{\pi q\lambda\beta\Bc}{2hc}\Big (1 +\frac{1}{\Bc\beta^2\gamma^2} \Big )
\tan \Big (\frac{\pi}{2}\bar{\delta}_0\Big ), \ \ \
G_{\rm eff} = \kappa\frac{4 q \lambda\beta\Bc}{h c} \Big (1 +\frac{1}{\Bc\beta^2\gamma^2} \Big ) \xi_1(\bar{\delta}_0). 
\label{s63-3}
\eeq

Image fields act as a perturbation on the betatron oscillation of the bunch.
Thus, the equation of motion of the perturbation $\Delta (s)$ of the transverse
coordinate of the bunch with the offset $\bar{\delta}= \bar{\delta_0}+ \Delta$ takes the form
\beq
{\Delta}^{''} -K_Q \Delta (s) = K_D,
\label{s63-4}
\eeq
with
\beq
K_D =  \frac{\pi}{2}\Ic\Big (1 + \frac{1}{\Bc\beta^2\gamma^2} \Big )
\tan \Big (\frac{\pi}{2}\bar{\delta}_0\Big ),\ \ \
K_Q = 4\Ic \Big (1 + \frac{1}{\Bc\beta^2\gamma^2} \Big ) \xi_1(\bar{\delta}_0).
\label{s63-5}
\eeq

As  the analysis in Ref. \cite{Wiedemann:2007yf}, Sect. 12.1.1 shows,
 dipole terms of the type $K_D$ cause a shift in the beam path 
 without affecting the focusing properties of the beam line.
In contract,  terms that depend linearly on the transverse bunch offset
from the orbit will affect focusing and the stability of the transverse motion of the beam,
because these perturbations act like quadrupoles.

The longitudinal size of the collimator $l_c$ is small in comparison with the wavelength 
of betatron oscillations, so we consider the collimator as a point source 
of perturbation. Let us assume that such a perturbation at point $(s_0, \bar{\delta}_0)$ is created by a 
thin lens of the focusing strength $\sigma_{\rm im} = g*l_c$ \cite{Wiedemann:2007yf, Pashkov:2006}. 
Here $g=qG_{\rm eff}/h m_p c\beta \gamma$, with $G_{\rm eff}$ from Eq. (\ref{s63-3}).

The evolution of a bunch through  elements of an accelerator is most clearly described 
by the product of transfer matrices \cite{Wiedemann:2007yf, Reiser:2008, Pashkov:2006}.
The transfer matrix for a full revolution is $\mathbf{M} =\mathbf{M_p}\mathbf{M_0}$, where for the lens
\beq
\mathbf{M_p} =
\left( 
\begin{array}{cc}
1 & 0 \\
\sigma_{\rm im} & 1 
\end{array} \right)
\eeq
and for the unperturbed part of the ring
\beq
\mathbf{M_0} =
\left( 
\begin{array}{cc}
\cos(\psi_0) & \hat{\beta_0}\sin(\psi_0) \\
-\frac{1}{\hat{\beta_0}}\sin(\psi_0) & \cos(\psi_0) 
\end{array} \right).
\label{s63-6}
\eeq
Here $\psi_0 = 2\pi\nu_0$ is the unperturbed phase advance per turn and $\hat{\beta_0}$ 
is the unperturbed
betatron function at the location of the perturbation, $s=s_0$.
On the other hand, the perturbed transfer matrix $\mathbf{M}$ can also be written  
in the form (\ref{s63-6}) if we replace
$\hat{\beta_0}$ by $\hat{\beta}$ and $\nu_0$ by  $\nu$, where $\nu$ is the the betatron 
tune in the presence of the  field gradient, $G_{\rm eff}$, from images.

\begin{figure*}[t]
\includegraphics[height=10.928cm,width=12.cm]{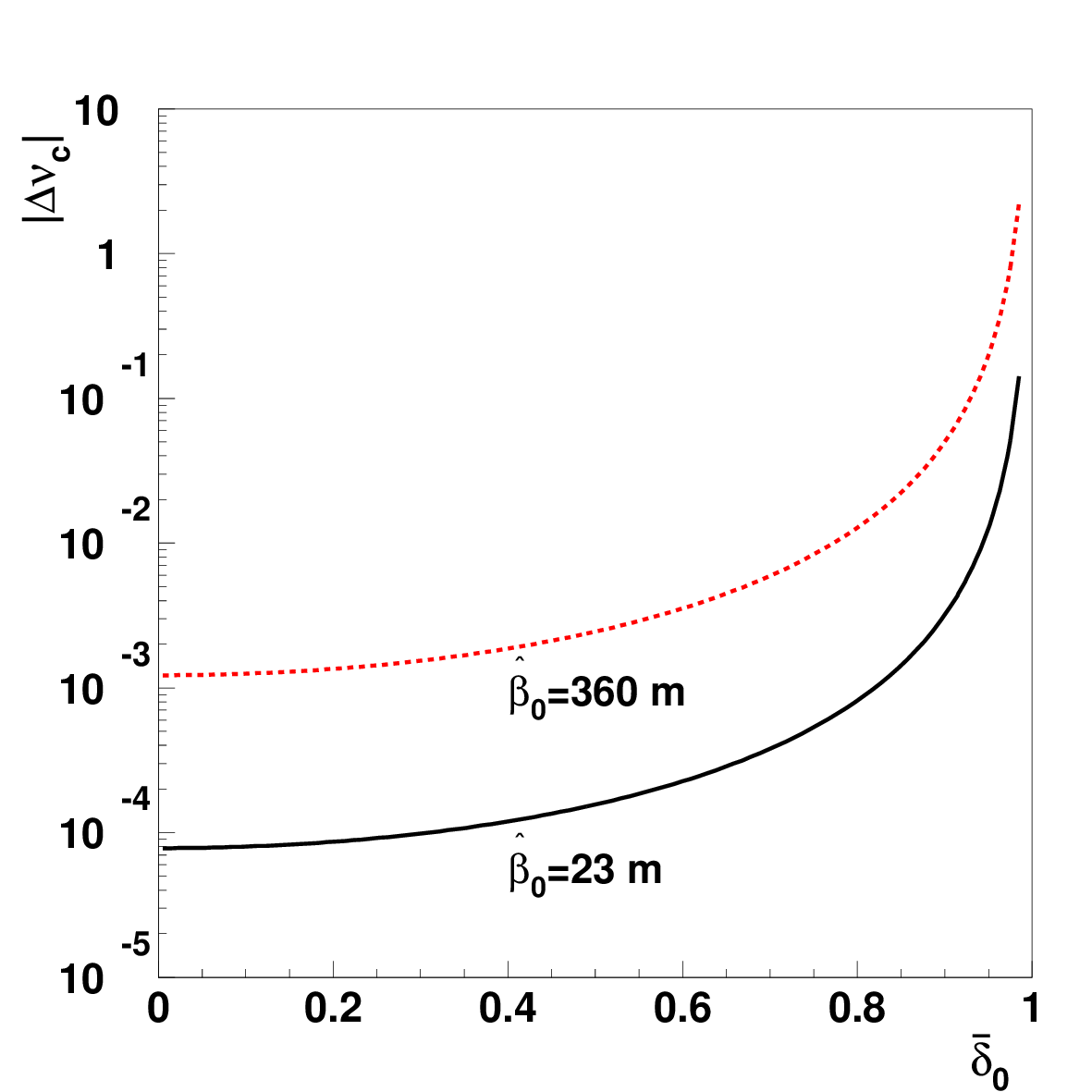}
\caption{
Variation of $|\Delta \nu_c(\bar{\delta}_0)|$ with the bunch offset at $\hat{\beta_0}=$23 m and 
$\hat{\beta_0}=$360 m.}
\label{fig7} 
\end{figure*}

By equating the traces of the perturbed matrix and the transfer matrix for a full revolution, 
$\Tr(\mathbf{M})=\Tr(\mathbf{M_pM_0})$, with setting $\nu=\nu_0+\Delta \nu$ and $|\Delta \nu| \ll \nu_0/2\pi$, we get the tune shift due to this perturbation,
\beq
\Delta \nu^{(\rm im)} (\bar{\delta}_0,h,l_c,\hat{\beta_0}) = -\frac{\hat{\beta_0}\sigma_{\rm im}}{4\pi} = -\frac{\hat{\beta_0}}{\pi}
\Ic\Big (1 + \frac{1}{\Bc\beta^2\gamma^2} \Big ) \xi_1(\bar{\delta}_0) \cdot l_c.
\label{s63-7}
\eeq 
The total tune shift is the sum of the individual contribution from each collimator
\beq
\Delta \nu^{(\rm im)}_{tot} = \sum_{i=1}^{n_c}\Delta \nu^{(\rm im)} (\bar{\delta_0}_i ,h_i,{l_c}_i,\hat{\beta_0}_i),
\eeq
where $n_c$ is the number of collimators along the accelerator  circumference.

As stated in Ref. \cite{Bruning:2004}, p. 100, the $\hat{\beta}$-functions at the collimators range from 27 m to 360 m
 and the range of collimator half-gaps $h$ is 4.7 to 11.1 mm for the injection optics 
and 1.2 to 3.8 mm for the squeezed optics.
On evaluating with Eq. (\ref{s63-7}) the  numerical values of  $\Delta \nu_c(\bar{\delta}_0)$, 
we set these values of $\hat{\beta_0}$, $h=1.2$ mm, $l_c=1$ m and the  machine and beam parameters from Table 1.
The results are presented in Fig.~\ref{fig7}.

Figure~\ref{fig7} shows that at  $ \bar{\delta}_0<0.6$
the total tune shift  
due to the quadrupole-type image fields in collimators,  
$\sim n_c\cdot|\Delta \nu^{\rm im}(\bar{\delta}_0)|$, $n_c  \sim 100$,
 does not exceed the safe value of the order of 0.1.
However, if the beam offset accidentally exceeds $\bar{\delta}_0 \approx 0.6\div 0.7$, it is possible to develop transverse instability and shift the betatron frequency into the resonance region.

%

\section{Summary}
This paper presents  new analytical expressions for electric and magnetic self-fields produced by a  bunch  shaped as a cylinder with a circular and an elliptical cross section. Calculations are done in the relativistic limit.
These expressions show the correct Coulomb asymptotic and in the near-field zone coincide with the external self-fields of a continuous beam.
 In the ultra-relativistic limit, external  fields of a bunch  takes  the universal form (\ref{s3-3}) and (\ref{s1-20}).

We reanalyzed the problem summation of image fields generated by a charged bunch between infinitely wide  parallel conducting plates and/or ferromagnetic poles. The  exact 1D solutions for resulting electric  and magnetic image fields are represented by the structure functions $\Lambda(\delta,\bar{\delta})$ and $H(\eta,\bar{\eta})$, respectively.

The new expressions for modified fields are applied to study  the coherent and incoherent tune shifts
for both infinite and finite parallel flat surfaces 
 and  allow  within an  improved linear approximation 
 generalization of the Laslett image coefficients  to the case of an arbitrary bunch offset 
$\bar{\delta}$.
These image coefficient functions,   $\epsilon_1(\bar{\delta})$ and $\xi_1(\bar{\delta})$, as well as  $\epsilon_2(\bar{\eta})$ and $\xi_2(\bar{\eta})$,   are not now independent but are rooted in the  functions $\Lambda(\delta, \bar{\delta})$ and $H(\eta,\bar{\eta})$,  correspondingly.
Equation (\ref{s63-7}) and Fig.~{\ref{fig7}} allow us to evaluate how the  gradient 
of image fields in a finite-size collimator affects the betatron frequency of the beam.
In particular, if the beam offset in collimators accidentally exceeds $\bar{\delta}_0 \approx 0.6\div 0.7$, it is possible to develop transverse instability and shift the betatron frequency into the resonance region. 

After a first version of the present paper became public \cite{Levchenko:2018a}
the author learned\footnote{Special thanks to the anonymous reader for providing Refs. \cite{Smythe:1950, Zotter:1975}  and useful comments.}
 about an old textbook  \cite{Smythe:1950} and  article \cite{Zotter:1975}, where only the scalar potential function of the electric field generated by a line charge between parallel earthed conducting planes is calculated with the use of conformal mapping. In addition, in \cite{Zotter:1975} are calculated incoherent and coherent image coefficients. 
The results of Ref. \cite{Zotter:1975} for the image coefficients and tune shifts
were rederived in Ref. \cite{Ng:2001} and several misprints are corrected.
This allows us to directly compare the results obtained by different methods.

Let us set, in  (3) from Sect. 4.20 of Ref. \cite{Smythe:1950}    $a=2h$, $b=h+\bar{y}$ and  $x=0$ (transition to a 1D problem). Then, by  expressing hyperbolic $\sinh$  via trigonometric functions $\sin$ and $\cos$, one exactly recovers Eq. (\ref{s4-12}).
Similarly, if one sets, in Eq. (27) of Ref. \cite{Zotter:1975}, $x=x_1=0$ and $y_1=\bar{y}$ and  applies
the double angle formula for  $\cos$, one gets Eq. (\ref{s4-12}). 
The field coefficients (29), (30) and (32) of Ref. \cite{Zotter:1975} 
 match Eqs.  (\ref{s4-10}), (\ref{s6-6-1}) and (\ref{s5-4}) of the present paper. 
However, expressions for the resulting electric and magnetic fields between the conducting plates (\ref{s4-8}), (\ref{s4-11}), (\ref{s5-8})
and the relations of type (\ref{s6-8}) between image coefficients and their origin were not revealed in Ref. \cite{Zotter:1975}.

The two parallel infinite plates is a particular case of a  rectangular vacuum chamber
when the width to height $w/h$ of the rectangle goes to infinity.
In Ref. \cite{Ng:2001} this limit was considered only for a centered beam, $\bar{\delta}=0$.
The electric and magnetic image coefficients as well as the tune shift  coefficients are of the same values as $\epsilon_{1,2}(0)$ and $\xi_{1,2}(0)$, above. The equations for
the vertical and horizontal, incoherent and coherent betatron tune shifts \cite{Ng:2001},
after some rearrangements and neglecting the neutralization factor, take the same form as  Eqs. (\ref{s6-7}), (\ref{s6-10}) and (\ref{s6-11}) at $\bar{\delta}= \bar{\eta}=0$.

Tune shifts and Laslett coefficients for a rectangular  and a circular beam pipe were also calculated 
in Refs. \cite{Petracca:1994, Petracca:1999}. Due to the fact that the radial and vertical betatron oscillations are coupled in general, the 
tune shift and image coefficients accordingly form  second-rank tensors. 
We can state a guess that the relations of type (\ref{s6-8}) are somehow connected with this fact; see Eq. (15) in Ref. \cite{Petracca:1994}. However, in Ref. \cite{Petracca:1994} explicit formulas have been given only for
a square vacuum chamber, $w=h$; therefore the passage to the limit of parallel plates is impossible.

To conclude, the results presented here can serve as a starting point in applying the  method of images to find exact solutions in  2D problems, calculating the field emission current in a collimator and a study of beam-induced multipacting. All these problems are relevant for the 
High-Limunosity Large Hadron Collider under construction \cite{CC2:2019}.

\section*{Acknowledgements}
The author is grateful to  P. Bussey, E. Lohrmann, M. Dohlus, and F. Willeke  
for  reading the early version of the manuscript, comments, and  useful discussions. 
I would like to express my special gratitude to E. Oborneva for her support 
and discussion of the results.

\appendix 

\renewcommand{\theequation}{A.\arabic{equation}}
\renewcommand{\thefigure}{A-\arabic{figure}}
\renewcommand{\thetable}{A-\arabic{table}}
\setcounter{equation}{0}  

\section{ Electric image  fields }\label{appA}

Here we derive the  formula (\ref{s4-4}).

Let us split the contribution of all image fields  (\ref{s4-1}) given in braces 
into two parts, 
\begin{eqnarray}
&  (2h-y_1)^{-1}\ -(2h+y_1)^{-1} 
- (4h-y_2)^{-1}\ +(4h+y_2)^{-1} \nonumber\\
&+  (6h-y_1)^{-1}\ -(6h+y_1)^{-1} 
- (8h-y_2)^{-1}\ +(8h+y_2)^{-1}  \nonumber\\
&+  (10h-y_1)^{-1}-(10h+y_1)^{-1}
- (12h-y_2)^{-1}+(12h+y_2)^{-1}+...\label{A1-1}\\
&= \sum_{\uk}^{\infty}\Pi^{(-)}_\uk (y_1,h) - \sum_{\um}^{\infty}\Pi^{(+)}_\um (y_2,h),
\label{A1-2}
\end{eqnarray}
where $\Pi^{(-)}_\uk$ represents the contribution from the  negatively charged images and
 $\Pi^{(+)}_\um$ is  the contribution from the positively charged images
\begin{eqnarray}
 \Pi^{(-)}_\uk (y_1,h)&=& \frac{1}{2\uk h-y_1} - \frac{1}{2\uk h+y_1}  
 = \frac{2}{h}\cdot\frac{\delta_1}{(2\uk)^2-\delta^2_1},
\label{A1-3}\\
 \Pi^{(+)}_\um (y_2,h)&=&\frac{1}{2\um h-y_2} - \frac{1}{2\um h+y_2}  
 =\frac{2}{h}\cdot\frac{\delta_2}{(2\um)^2-\delta^2_2} .
\label{A1-4}
\end{eqnarray}
Here and hereinafter,  indexes $\uk$ and $\um$ have odd, $\uk\,=\,$1,3,5,..., and even, 
$\um\,=\,$2,4,6,... values, $\delta_1=y_1/h$ and $\delta_2=y_2/h$.

Now it is  evident that the space structure of the image fields between plates is 
described by a specific function $\Lambda (\delta_1,\delta_2)$  we  term this  
the structure function:
\beq
\sum_{\uk}^{\infty}\Pi^{(-)}_\uk - \sum_{\um}^{\infty}\Pi^{(+)}_\um=\frac{2}{h}\Lambda (\delta_1,\delta_2).
\label{A1-5}
\eeq
with
\beq
\Lambda (\delta_1,\delta_2)=\delta_1\sum_{\uk}^{\infty}\frac{1}{(2\uk)^2-\delta^2_1}
-\delta_2\sum_{\um}^{\infty}\frac{1}{(2\um)^2-\delta^2_2}.
\label{A1-6}
\eeq
The structure function $\Lambda$  depends only on the  scaled variables.
 
To proceed further, recall the decompositions  (1.421) \cite{gradshteyn:1971} 
\begin{eqnarray}
 \tan(\frac{\pi}{2}z)&=&\frac{4}{\pi}z\sum_{\un=1}^{\infty}\frac{1}{(2\un-1)^2-z^2},\nonumber\\
  \cot(\pi z)&=&\frac{1}{\pi z} -\frac{2z}{\pi}\sum_{\un=1}^{\infty}\frac{1}{\un^2-z^2}.
\label{A1-7}
\end{eqnarray}
After some algebraic manipulation and the use of Eq. (\ref{A1-7}), we get from Eq. (\ref{A1-6}) a new exact and compact expression of the structure function
\beq\
\Lambda (\delta_1,\delta_2)=\frac{1}{2} \Big [ \frac{\pi}{4}\tan ( \frac{\pi}{4}\delta_1)\, +\,
\frac{\pi}{4}\cot (\frac{\pi}{4}\delta_2)\,-\,\frac{1}{\delta_2}\Big ].
\label{A1-8}
\eeq
Now, if  we recall that $\delta_1 = (y+\bar{y})/h = \delta +\bar{\delta} $ and 
$\delta_2 = (y - \bar{y})/h = \delta -\bar{\delta} $, we obtain
\beq 
\Lambda (\delta,\bar{\delta})=\frac{1}{2} \Big [\frac{\pi}{2}\cdot \frac{\cos(\frac{\pi}{2}\bar{\delta})}
{\sin(\frac{\pi}{2}\delta)-\sin(\frac{\pi}{2}\bar{\delta})} -\frac{1}{\delta -\bar{\delta}}\Big ].
\label{A1-9}
\eeq

For some applications it is more practical to use the relations between the Bernoulli numbers and 
the trigonometric functions. To do this, recall the decompositions (1.411) \cite{gradshteyn:1971}
\begin{eqnarray}
 z\cdot\tan(z) &=& \sum_{\un=1}^{\infty}\frac{(2^{2\un}-1)(2z)^{2\un}}{(2\un)!}|B_{2\un}|,
\nonumber \\
 z\cdot\cot(z) &=&  1-\sum_{\un=1}^{\infty}\frac{(2z)^{2\un}}{(2\un)!}|B_{2\un}|.
\label{A1-10}
\end{eqnarray}
where $B_{2\un}$ are Bernoulli numbers, $B_2=1/6$, $B_4=-1/30$, $B_6=1/42$ etc.
After substituting Eq. (\ref{A1-10}) in Eq. (\ref{A1-8}), we find the following form of 
the structure function 
\beq 
\Lambda (\delta_1,\delta_2)=\frac{1}{2}\sum_{\un=1}^{\infty}
\Big [ (2^{2\un}-1)\delta_1^{2\un-1} - \delta_2^{2\un-1}\Big ] 
\frac{\pi^{2\un}}{2^{2\un}(2\un)!}|B_{2\un}|\,.
\label{A1-11}
\eeq
Using only the linear terms,  we recover the  part derived by Laslett \cite{Laslett:1963jn}
 (see Eq. (\ref{s4-2}))
\beq 
\Lambda (\delta,\bar{\delta})= \frac{1}{h}\cdot\epsilon_1 (y+2\bar{y}).
\label{A1-12}
\eeq 
An inspection of Eq. (\ref{A1-11}) shows that the contributions of negatively  charged images
are enhanced by the factor $2^{2n}-1$, as compared with the contributions from  the positively 
charged images. Equation (\ref{A1-11}) also shows that for $y$ on the bunch axis, i.e. $\delta_2 =0$, 
 the contributions from the  positively charged images  vanish.

At  first glance,  Eqs. (\ref{A1-8}) or (\ref{A1-9}) are singular at $\delta_2=0$ or
$\delta=\bar{\delta}$, respectively. However,   this is not the case. 
Starting once again from Eq. (\ref{A1-11})  with $\delta_2=0$ and taking  
Eq. (\ref{A1-10}) into account, we formally get 
\beq 
\Lambda(\bar{\delta},\bar{\delta})=\frac{\pi}{8}\tan \big (\frac{\pi}{2}\bar{\delta}\big ).
\label{A1-13}
\eeq

Knowing the exact form of $\Lambda (\delta,\bar{\delta})$, one is able to develop a variety of approximation. 
For instance, to study particle dynamics in a bunch with a significant offset, one needs 
to decompose Eq. (\ref{A1-8}) assuming $\delta_2 \ll 1$. The result is
\beq
 \Lambda(\bar{\delta},\delta_2) \simeq \Lambda(\delta_1,\delta_2)_{\mid\delta_2=0}
+{\frac{\partial \Lambda}{\partial \delta_2}{\Big \vert}}_{\delta_2=0}\!\!\!\!\cdot \delta_2
 =\frac{\pi}{8}\tan(\frac{\pi}{2}\bar{\delta}) +\frac{\pi^2}{32}
\Big [ \frac{1}{\cos^2(\frac{\pi}{2}\bar{\delta})} - \frac{1}{3}\Big ]\delta_2 .
\label{A1-14}
\eeq
%


\renewcommand{\theequation}{B.\arabic{equation}}
\renewcommand{\thefigure}{B-\arabic{figure}}
\renewcommand{\thetable}{B-\arabic{table}}
\setcounter{equation}{0}  

\section{Magnetic image fields } \label{appB} 

Let the boundary of magnet pole faces  be represented as two parallel plates located at
$y=\pm g$.  A magnetic field, seen by a particle at location $y$ on the $y$-axis, is generated by
the successive image currents with the same sign as the beam itself 
\cite{Chao:1993zn, Hofmann:1992ki}. Therefore, instead of the series (\ref{A1-1})    
we get
\begin{eqnarray}
&&  (2g-y_1)^{-1}\ -(2g+y_1)^{-1}
+(4g-y_2)^{-1}\ -(4g+y_2)^{-1} \nonumber\\ 
&+&  (6g-y_1)^{-1}\ -(6g+y_1)^{-1}
+(8g-y_2)^{-1}\ -(8g+y_2)^{-1}\nonumber\\ 
&+&  (10g-y_1)^{-1}-(10g+y_1)^{-1}
+(12g-y_2)^{-1}-(12g+y_2)^{-1}+... \\
\label{C-1} 
&=& \sum_{\uk}^{\infty}\Pi_\uk (y_1,g) + \sum_{\um}^{\infty}\Pi_\um (y_2,g)\nonumber\\ 
&=& \frac{2}{g} H(\eta_1,\eta_2)
\label{C-2}
\end{eqnarray}
where  $\Pi_\uk$ and $\Pi_\um$ are of the same functional form as Eqs. (\ref{A1-3}) and (\ref{A1-4})  
with an interchange of variables   $h\rightarrow g$, $\delta_1 \rightarrow \eta_1=y_1/g$, 
$\delta_2 \rightarrow \eta_2=y_2/g$.
Indexes $\uk$ and $\um$ have odd, $\uk\,=\,$1,3,5,..., and even,  $\um\,=\,$2,4,6,...  
values, respectively.

Now, if we are to proceed in the same manner as in Appendix \ref{appA}, we obtain from Eq. (\ref{C-1})   
 an expression for the structure function  of the image magnetic fields 
\beq 
H(\eta_1,\eta_2)=\frac{1}{2}\sum_{\un=1}^{\infty}\Big [ 
(2^{2\un}-1)\eta_1^{2\un-1} + \eta_2^{2\un-1}\Big ]
\frac{\pi^{2\un}}{2^{2\un}(2\un)!}|B_{2\un}|\, .
\label{C-3}
\eeq
For $|y|\ll g$ and $|\bar{y}|\ll g$, this gives, to first order in $\eta_1$ and $\eta_2$
\beq
H(y,\bar{y})= \frac{\epsilon_2}{g}\cdot (y+\frac{1}{2}\bar{y})
=\epsilon_2(\eta + \frac{1}{2}\bar{\eta}).
\label{C-4}
\eeq  
The use of Eq. (\ref{A1-10}) in Eq. (\ref{C-3})  gives  the magnetic
structure function that we are looking for
\beq 
H(\eta_1,\eta_2)=\frac{1}{2} \Big [ \frac{\pi}{4}\tan ( \frac{\pi}{4}\eta_1)\, -\,
\frac{\pi}{4}\cot (\frac{\pi}{4}\eta_2)\,+\,\frac{1}{\eta_2}\Big ].
\label{C-5}
\eeq
Now, if  we recall the relationships $\eta_1 = (y+\bar{y})/g = \eta +\bar{\eta} $ and 
$\eta_2 = (y - \bar{y})/h =  \eta -\bar{\eta} $, we obtain
\beq
H(\eta,\bar{\eta})\,=\,\frac{1}{2}\Big [\frac{1}{\eta -\bar{\eta}} -\frac{\pi}{2}\cdot
\frac{\cos(\frac{\pi}{2}\eta)}{\sin(\frac{\pi}{2}\eta)-\sin(\frac{\pi}{2}\bar{\eta})}\Big ].
\label{C-6}
\eeq

With the same reasoning as in  Appendix \ref{appA},  
the full linear approximation in $ \eta_2$ is given by
\beq 
H(\eta,\bar{\eta})\simeq  \frac{\pi}{8}\tan ( \frac{\pi}{2}\bar{\eta})\, +\,
\epsilon_2 (\bar{\eta})(\eta - \bar{\eta}),
\label{C-7}
\eeq
where $\epsilon_2 (\bar{\eta})$  is defined in Eq. (\ref{s5-4}).


\renewcommand{\theequation}{C.\arabic{equation}}
\renewcommand{\thefigure}{C-\arabic{figure}}
\renewcommand{\thetable}{C-\arabic{table}}
\setcounter{equation}{0}  

\section{ ``The principle of electric images'' }\label{appC}

Throughout his long and fruitful scientific life, William Thomson (later Lord Kelvin) was in active correspondence with his father,  the well known  mathematician James Thomson as well as 
 with leading scientists in Great Britain and Europe.
The topics of the letters were both personal and discussion of the latest scientific 
news and emerging new ideas. In addition, at the insistence of his father, W. Thomson 
kept diaries in mathematics and physics. These letters and diaries have been preserved and published \cite{Wilson:1990, Thompson:1910},  allowing us  to trace the roots of ideas and the time of their origin.                             

As follows from the mathematical diary of 21-year-old W. Thomson, 
the story of  ``the principle of electric images'' began in 1845, during his 
four-month stay in Paris.  This trip abroad  had several objectives. 
First, to improve the physical condition
and also personally get acquainted with the leading 
 French mathematicians (Liouville, Cauchy, Sturm), 
physicists (Arago, Biot, Pouillet), chemists (Regnault) 
and attend their lectures at the  {\'E}cole Polytechnique 
and  the Coll\`ege de France \cite{Thompson:1910}.
Before  Thomson had been a month in Paris he sent to Liouville's {\it Journal de Math\'ematiques}
(vol. X. p. 137, 1845) a paper, entitled ``D\'emonstration
d'un th\'eor\`eme d'analyse'', which two years later
he expanded in the {\it Cambridge and Dublin Mathematical Journal} (vol. II. p. 109, March 1847)
under the title: ``On certain definite integrals suggested by problems in the theory of electricity''. 
This article allowed Thomson to demonstrate his skills as a mathematician
and identified his area of interest as a natural philosopher.

During long conversations and discussions of the results of Faraday's experiments 
and problems in the development of the mathematical theory of electricity, 
J. Liouville engaged W. Thomson to write a memoir for the Institute, 
with his vision of their solution \cite{Thompson:1910}.
Common interests in mathematics and physics made friends Thomson and Liouville.
Liouville's friendship meant much to Thomson. To  Liouville he confided his ideas 
about electric images. 

From  the mathematical diary (\cite{Thompson:1910}, p. 126):
``March 15, 1845--I am occupied the whole day in Regnault's 
physical laboratory at the Coll\`ege de France.
At spare times I have been reading Poisson's memoirs
on electricity, which I find among the Memoirs of the
Institute in Regnault's cabinet. I have applied my ideas on induction in spheres 
and the principle of successive influence, and get a very simple 
solution, in the form Poisson gives it, for two spheres. I think I can
work it out for $i$ spheres ... The {\it image} of an exterior
point, in a conducting sphere, is a p. in the interior, with
opposite electr.''

In a footnote added in 1872  to one of his 1848 articles, Thomson describes this period as follows 
(Ref. \cite{Thomson:1884}, p. 52):
\begin{quote}
``... A complete exposition of the principle of electrical images (of which a short
account was read at the late meeting of the British Association at Oxford) has
not yet been published; but an outline of it was communicated by me to
M. Liouville in three letters, of which extracts are published in the Journal de
Math\'ematiques (1845 and 1847, vols. X., XII.).  A full and
elegant exposition of the method indicated, together with some highly interesting
applications to problems in geometry not contemplated by me, are given by
M. Liouville himself, in an article written with reference to those letters, and
published along with the last of them. I cannot neglect the present opportunity
of expressing my thanks for the honour which has thus been conferred upon me
by so distinguished a mathematician, as well as for the kind manner in which
he received those communications, imperfect as they were, and for the favourable
mention made of them in his own valuable memoir.''
\end{quote}

Thomson does not seem to have made the acquaintance of G.~G. Stokes till
after his return from Paris in May 1845; at least Stokes's name 
never occurs in Thomson's earlier letters or in his diary. 
Yet he was settled in Pembroke as a junior fellow, having been Senior
Wrangler in 1841. It seems that it was the task
of editing the {\it Cambridge and Dublin Mathematical Journal} that brought
them together; and Stokes's penchant for experimenting led Thomson often to seek his advice.
Stokes was indeed guide, philosopher, and friend
to his eager and enthusiastic disciple.
Many years later Lord Kelvin said (\cite{Stokes_memo_v1}, p. 318):
 ``For sixty years of my own life, from 1843 to 1903, I looked up
to Stokes as my teacher, guide, and friend.'' 
For more than fifty years (1846--1903) each was in the
habit of communicating to the other the progress of
his ideas.

On June 20, 1847,  Thomson writes from Peterhouse to his
father about the approaching meeting of the British
Association (Ref. \cite{Thompson:1910}, p. 204):
``...I have been getting out various interesting pieces
of work, along with Stokes, connected with some problems
in electricity, fluid motion, etc., that I have been thinking
on for years, and I am now seeing my way better than I
could ever have done by myself, or with any other person
than Stokes.''

In July 1847, Thomson and Stokes attended the meeting of
the British Association at Oxford with talks entitled
``On Electrical Images'' (by W. Thomson) and
``On the Resistance of a Fluid to two Oscillating Spheres'' (by G.~G. Stokes).  
One year later the Report of the Seventeenth Meeting of the British Association
was published; however, the results of Thomson  and Stokes
were presented by short abstracts. 
Some  extractions from these abstracts follow.

\noindent W. Thomson \cite{Thomson_Rep_BA_1847}:
\begin{quote}
``There is no branch of natural philosophy of which the elementary laws are more
simple than those which regulate the distribution of electricity upon conducting
bodies; yet its impracticability has always been the reproach of the mathematical
theory of electricity. Very few of the varied and interesting problems which it 
presents have been made subjects for investigation, on account of the apparently extreme
complexity of the conditions to be satisfied; and even when results have been forced
from it by the analytical skill and energy of a Poisson, the physical interest has been
almost lost in the struggle with mathematical difficulties, and the complexity of the
solution has eluded that full interpretation without which the mind cannot be 
satisfied in any analytical operations having for their object the investigation or 
expression of truth in natural philosophy...

The subject of this communication is `the principle of electrical images,' which
is suggested by Green's elementary propositions, as the proper way of treating a
great variety of problems that present themselves with reference to the 
distribution of electricity on spherical conductors. The effect of a body electrified 
in any given manner upon an uninsulated sphere is shown to be completely represented
by what may be called ``the image of the electrified body in the sphere,'' and 
a simple geometrical construction is given by which this image may be described. 
When an electrified body is placed in the neighbourhood of two uninsulated spheres, 
an inductive effect is produced which may be represented by an infinite series of 
`successive images' in each sphere. An algebraic expression of this result leads 
to solutions, by means of converging series, of the various problems which occur with 
reference to the distribution of the induced electricity, and the attractions exerted by
the two spheres. When a single conductor bounded by segments of two spherical
surfaces cutting at an angle which is a submultiple of two right angles is electrified
by the influence of a charged body, the effect may be represented by means of a finite
number of images disposed in a symmetrical manner in the circumference of a circle
passing through the exciting body, and cutting the two spherical surfaces at right
angles. The principle of electrical images, as applied in these two cases, may be
illustrated by a reference to the successive images of a candle placed between two
parallel plane mirrors, and to the symmetrically arranged images which are seen in
the kaleidoscope.''
\end{quote}

\noindent G. G. Stokes \cite{Stokes_Rep_BA_1847}:
\begin{quote}
``The object of this communication was to shew the application
of Professor Thomson's method of images to the solution of certain
problems in hydrodynamics...

The investigation mentioned in the preceding paper arose out
of the communication to me by Sir William Thomson of his
beautiful method of electrical images before he had published it.
Having myself paid more attention to the motion of fluids than
to electricity, I endeavoured to find if it would in any manner
apply to the solution of problems in the motion of fluids. I found
that what is called above a singular point  of the second 
order \footnote{Currently called a doublet or dipole.}
 had a perfect image in a sphere when its axis was in the 
direction of a radius, which led to a complete solution of the problem 
mentioned in the paper when one sphere lay wholly outside or inside
the other.''
\end{quote}

Many years later, an extended version of this report was published in Ref. \cite{Stokes:1880}.
After this article, the method of images became one of the problem-solving techniques 
in hydrodynamics \cite{Lamb:1932, Newman:1999}.

It should be noted that the technique of electrical images is described by Thomson himself fragmentary and scattered on the articles of different years. The most consistent presentation of this method with use of
the idea of the potential and of equipotential surfaces is given in 
``A Treatise on Electricity and Magnetism'' by Maxwell \cite{Maxwell:1873a}.
This is how Maxwell begins to introduce the reader to the principle of image charges:
\begin{quote}
``... In applying this method to the most elementary case of a sphere
under the influence of a single electrified point, we require to expand
the potential due to the electrified point in a series of solid 
harmonics, and to determine a second series of solid harmonics which
express the potential, due to the electrification of the sphere, in the
space outside. It does not appear that any of these mathematicians observed
that this second series expresses the potential due to an imaginary
electrified point, which has no physical existence as an electrified
point, but which may be called an electrical image, because the
action of the surface on external points is the same as that which
would be produced by the imaginary electrified point if the spherical
surface were removed.

This discovery seems to have been reserved for Sir W. Thomson,
who has developed it into a method of great power for the solution
of electrical problems, and at the same time capable 
of being presented in an elementary geometrical form...''  \cite{Maxwell:1873a}
\end{quote}

Thus, the method of image charges and currents allows us to solve problems on the distribution of real charges and fields not only due to the conductive surface of the simplest geometric form, a plane, but also for conductors of more  complex  geometric shapes.

\bibliographystyle{./ptephy}
\bibliography{levchenko20}

\end{document}